\documentclass[12pt]{article}
\usepackage{geometry}                
\usepackage{graphicx}
\usepackage{amssymb}
\usepackage{epstopdf}
\usepackage{color}
\usepackage{setspace}
\title{Systematic fluctuation expansion for  neural network activity equations.}
\author{Michael A. Buice\footnote{buicem@niddk.nih.gov}\\ {\small Laboratory of Biological Modeling/NIDDK/NIH, Bethesda, MD, 20892}
\\Jack D. Cowan\footnote{cowan@math.uchicago.edu}\\{\small University of Chicago, Depts of Mathematics and Neurology, Chicago, IL, 60637}\\
Carson C. Chow\footnote{carsonc@mail.nih.gov}\\{\small Laboratory of Biological Modeling/NIDDK/NIH, Bethesda, MD, 20892}}
\begin{document}
\maketitle
\abstract{Population rate or activity equations are the foundation of a common approach to modeling for neural networks.  These equations provide mean field dynamics for the firing rate or activity of neurons within a network given some connectivity.  The shortcoming of these equations is that they take into account only the average firing rate while leaving out higher order statistics like correlations between firing.  A stochastic theory of neural networks which includes statistics at all orders was recently formulated.  We describe how this theory yields a systematic extension to population rate equations by introducing equations for correlations and appropriate coupling terms.  Each level of the approximation yields closed equations, i.e. they depend only upon the mean and specific correlations of interest, without an {\it ad hoc} criterion for doing so.  We show in an example of an all-to-all connected network how our system of generalized activity equations captures phenomena missed by the mean field rate equations alone.}

\section{Introduction}

Modeling the brain is confounded by the fact that there are a very large number of neurons and the neurons are heterogeneous and individually complex.  Given current analytical and computational capabilities, we can either study neuronal dynamics in some biophysical detail for a small or medium set of neurons or consider a large population of abstract simplified neural units.  We then can only extrapolate to the desired regime of large numbers of biophysical neurons.  In particular, there is a dichotomy between network models that incorporate Hodgkin-Huxley or integrate-and-fire spiking dynamics and models that only include the rate or activity of neural units.  While the rate description has yielded valuable insights into many neural phenomena it cannot describe physiological phenomena thought to be important for neural processing such as synchronization, spike-time dependent plasticity or any correlated activity at the spike level.  Likewise, it is difficult to analyze or simulate a large network of spiking neurons.  Our goal is to derive an intermediate description of neural activity that is complex enough to account for spike level correlations yet simple enough to be amenable to analysis and numerical computation for large networks.

Rate or activity equations have been a standard tool of computational and theoretical neuroscience, early important examples being the work of Wilson and Cowan, Cohen and Grossberg, Amari and Hopfield~\cite{wilson1,wilson2,amari1,amari2,hopfield2,cohen}.  Models of this type have been used to investigate pattern formation, visual hallucinations, content addressable memory and many other questions~\cite{ ermentrout, hopfield2,bard1,bressloff,coombes1}. Naturally, these equations are so called because they describe the evolution of a neural activity variable often ascribed to the firing rate or synaptic drive of a population of interacting  neurons \cite{bard1,gerstner1}.  These equations are considered to represent the neural dynamics averaged over time or population of a more complicated underlying process.  In general, these activity equations make an implicit assumption that correlated firing is unimportant.   They are a ``mean field theory" which capture the dynamics of the mean firing rate or activity that is independent of the influence of correlations, which in some cases may alter the dynamics considerably.  As an example, the effects of synchrony, which have been proposed to be important for neural processing \cite{CMGray031989,kopell} are not included.  Here, we give a systematic prescription to extend rate models to account for these effects.

An  analogy for our problem and approach can be made to the field of equilibrium statistical mechanics.  The statistics of such systems (e.g. the Ising model) in thermal equilibrium are described by a partition function, which is an integration over all configurations available to the system.  For the Ising model this refers to  all possible configurations of the individual spins.  The partition function is akin to the generating function for a statistical distribution from which the moments or cumulants can be obtained.  For the Ising model the first moment corresponds to the mean magnetization and the second moment describes the mean correlation between the spins.  The linear response of the system is the magnetic susceptibility, which describes the reaction of the system to an external input.  In general the partition function cannot be summed or integrated explicitly.  However, these moments can be obtained perturbatively by using the method of steepest descents to approximate the partition function.  This then yields a systematic expansion and the lowest order is called mean field theory, since all higher cumulants are zero.  By computing the expansion to higher order, the effects of correlations and fluctuations can be included.

This procedure requires full knowledge of the underlying microscopic theory that is to be averaged over.  In neuroscience, the underlying model is not completely  known; it would require full knowledge of the different types of neurons, their membrane and synaptic kinetics, and their synaptic connectivity.  However, given a particular mean field theory, one can ask about the minimal constraints this theory places upon the microscopic theory and its asymptotic expansion.  Thus, although the full microscopic theory cannot be reconstructed, by constraining the expansion, the mean field theory can dictate the minimal structure of any extension of a set of rate equations.  In this paper, we consider a well known neural rate equation and deduce the minimal structure we expect for a consistent extension which includes correlations.

Buice and Cowan~\cite{bc} previously adapted a path integral formalism used in nonequilibrium statistical mechanics \cite{doi1,doi2,peliti} to analyze the  dynamics of a Markov model for neural firing. They derived a generating functional (expressed as an infinite dimensional path integral), which is specified by an ``action" for the complete dynamical distribution of the model and showed  that the mean field theory for that system corresponded to a Wilson-Cowan-type rate equation.  They then analyzed the scaling properties  for the correlations near criticality.   They also showed how mean field theory could be corrected by using steepest descents to generate a systematic expansion that describes the effect of correlations.  
Here, we show that by taking explicit averages of the Markov model a moment hierarchy can be constructed.  Each equation in the moment hierarchy is coupled to higher moments in the hierarchy.  The hierarchy can be made useful as a calculational tool for the statistics of the dynamics if it can be truncated.  We show that the moment hierarchy and the generating functional are equivalent and that the equations of the hierarchy are the ``equations of motion" of the action in the generating functional.   The truncation condition for the perturbation series of the path integral is also a truncation condition for the hierarchy.  This provides for both a compact description of network statistics and a natural truncation or closure condition for a moment hierarchy.   We can also show using the path integral formalism that the Markov model is a natural minimal extension of the Wilson-Cowan rate equation.

Approaches to neural network modeling using statistical mechanics are not new \cite{hopfield1,hopfield2,peretto,somp3}.  Those works were largely concerned with models adhering to detailed balance, whereas we make the explicit assumption that neural dynamics admits an absorbing state that violates detailed balance.  In the absence of internal activity and external stimulation, there will be no activity in the network.  Other studies using a stochastic description of neural dynamics have considered the neurons in a background of Poisson activity with disorder in the connectivity \cite{DJAmit04011997, amitbrunel1}, or considered neural activity as a renewal process \cite{PhysRevE.51.738, gerstner1, gerstner2}. Van Vreeswijk and Sompolinsky  [1996,1998] 
demonstrated that disorder in network activity can arise purely as a result of disorder in the connectivity, without stochastic input.  Kinetic theory and density approaches are investigated in \cite{Nykamp00apopulation,DavidCai05182004,  Ly:2007p11} and mean field density approaches to the asynchronous state appear in \cite{PhysRevE.48.1483,Treves}.  \cite{golomb1} study synchrony in sparse networks via a reduction to a phase model.   Fokker-Planck approaches for networks appear in \cite{fusi1, hakim, brunel2,  brunel1}.  Responses of single neurons driven by noise appear in \cite{plesser, salinas2, fourcaud, soula}.  
Approaches  to correlated neural activity including finite size effects appear in \cite{somp1,PhysRevE.66.051917,soula2,dest}.  
In \cite{dest}, the authors develop a moment hierarchy for a Markov model of asynchronous irregular states of neural networks which is truncated through a combination of finite size and a scaling condition.  Our work extends the results of \cite{somp1} by providing the systematic higher order expansion without explicitly requiring the consideration of the rest of the hierarchy.  We also provide conditions for the truncation of the expansion and consider the network response to correlated input.  Our expansion is not a finite size expansion, although it can reduce to a finite size expansion under certain conditions (such as normalized all-to-all connectivity in the network).  

In section~\ref{sec:wcredux}, we revisit the original Wilson-Cowan framework and propose a Markov model that has the minimal stochastic dynamics to produce the Wilson-Cowan equations.  This will be more rigorously justified in section~\ref{sec:field}.  Section~\ref{sec:momentHierarchy} presents the derivation of a moment hierarchy for this Markov model.  After truncating, we provide {\it a posteriori} justification for the truncation.  It will be seen in section~\ref{sec:field} that the validity of this truncation was in fact natural and did not require {\it ad hoc} assumptions.  The truncation conditions turn out to be related to the proximity to a bifurcation point as well as the extent of connectivity in the network.   We also make more precise the sense in which our Markov model is ``minimal" by introducing the path integral formulation.   The field theory formalism which appears in this paper arose in the context of reaction-diffusion problems.  See \cite{Janssen2005147,0305-4470-38-17-R01} for reviews of this formalism  applied to reaction-diffusion and percolation processes.  We demonstrate a simple example all-to-all system in Section~\ref{sec:alltoall} and show some simulation results. 


\section{Rate equations reconsidered}
\label{sec:wcredux}
We consider a population rate equation of the form
\begin{equation}
	\partial_t a({x},t) = - \alpha a({x},t) + f \left ( \int d^d{y} \,w({x} , {y}) a({y},t) + I(x,t) \right) 
	\label{eq:wc}
\end{equation}
where $a(x,t)$ is a measure of local neural ``activity" at location $x\in {\mathbb R}^d$, $\alpha$ is a decay constant (often equated with the membrane time constant or a synaptic time constant), $f(s)$ is the firing rate or gain function describing how input affects the activity,  $I(x,t)$ is a time dependent external input to location $x$, and $w({x},{y})$ is a weight function describing how a neuron at location ${y}$   affects neurons at location ${x}$.  Equation~(\ref{eq:wc}) is the standard form of rate equation seen in the Wilson-Cowan equations \cite{wilson1,wilson2}. While we use this form in our paper,  our results can be adapted to any other type of rate or activity equation.
The exact nature of the activity $a(x,t)$ is open to interpretation \cite{PhysRevE.51.738,bard1}.  It could be envisioned as the time average, population average or ensemble average of neural firing or synaptic activity.  In any case, the picture is that of some underlying process whose degrees of freedom have been marginalized to generate an effective theory with simpler variables.

We imagine that the typical rate equation is produced by some marginalization process over both disorder and extra degrees of freedom. Hence, it may be possible to derive a generating function for the statistics of the marginalized process.  The lowest order in the steepest descent expansion of the generating function describes ``mean field theory", which gives the rate equation.    Since the operation of marginalization is dissipative, we cannot recreate the underlying microscopic process exactly with only the rate equation alone.  However, the mean field theory places constraints upon the structure of the dynamics, enabling us to investigate the structure of higher order statistics implied by the structure of mean field theory.  In  the original derivation by Wilson and Cowan~ \cite{wilson1,wilson2}, 
 the activity variable was presumed to describe the fraction of neurons firing per unit time  within some region of the brain.  There are two main features of this interpretation which bear emphasizing.  First, the rate equations were originally understood to be equations providing the dynamics of the probability that a neuron at $x$ will fire at time $t$.  There is therefore an implied underlying  \emph{probabilistic} model.  Second, the probability $a(x,t)$ applies to \emph{all} neurons within some region of the brain, not just a single neuron.  Thus, there is a spatial averaging component implicit in the equations.  The original Wilson-Cowan rate equations thus described the dynamics of the probability for a neural aggregate in the brain.  Another feature implicit in the Wilson-Cowan equations is that these probabilities are independent for each neuron.  This implies that the Wilson-Cowan picture is one in which neurons fire with Poisson statistics with firing rate determined by $a(x,t)$, a picture supported by neural recordings \cite{softky}.


Given this perspective, one might consider what processes may underly rate equations.  One route is to treat the fundamental, small-scale dynamics as a probabilistic process, for example a Markov process.  In this case, the basic description for neural activity will be provided by a master equation governing the evolution of probabilities for different neural configurations.  This route obscures the source of uncertainty in neural activity in favor of directly modeling the probabilistic activity.  This tactic has been used to model the so-called asynchronous irregular states seen in some neural models \cite{Vreeswijk96chaosin,dest}.  Another route would be to employ the strategy of kinetic theory \cite{nicholson,ichimaru} and define a continuity  (i.e. Klimontovich) equation for the probability density of a network of deterministic neurons.  For an example of this approach applied to coupled oscillators, see \cite{Hildebrand:2006p4,buice:031118}.  In that work, the probabilistic aspects of the model arise from the distribution of driving frequencies and initial conditions.
Ultimately, the difference in the two approaches is the origin of stochasticity, i.e. whether it is implicit in the dynamics  of the neurons or an emergent property of the interaction of deterministic neurons (e.g. chaos).  In either case, the final product is an effective stochastic dynamical system.  In this paper, we will follow the approach of assuming an underlying probabilistic model given by a master equation, so that any emergent chaos  has already been absorbed into the dynamics.  We then seek a minimal stochastic model that will produce the Wilson-Cowan rate equation at the mean field level.  We can then formulate equations governing the fluctuations of this model.  In this section we motivate such a minimal model qualitatively, leaving a more rigorous approach for section~\ref{sec:field}.

 Our primary interest is in tracking the statistics of active neurons.   A simple master equation whose mean field statistics for neural activity is represented by the Wilson-Cowan rate equations is given by
 \begin{eqnarray}
	\frac{d P(\vec{n}, t)}{dt} &=& \sum_i \left [ \alpha (n_i + 1) P(\vec{n}_{i+},t) -  \alpha n_iP(\vec{n},t)  \right .\nonumber \\
	&&+  \left . F_i\left (  \vec{n}_{i-} \right) P(\vec{n}_{i-}, t)  - F_i\left (  \vec{n} \right) P(\vec{n}, t)  \right ]
	\label{eq:master}
\end{eqnarray}
where $P(\vec{n}, t)$ is the probability of the network having the configuration described by $\vec{n}=\{n_1,n_2,\cdots\}$ at time $t$, and $n_i$ is the number of active neurons at location $i$.  Neurons relax back to the inactive or quiescent state with rate $\alpha$, which appears as a decaying transition in the master equation.  Configurations $\vec{n}_{i+}$ and $\vec{n}_{i-}$ denote the configuration $\vec{n}$ where the $i$th component is $n_i\pm1$, respectively. The rate at which a neuron at location $i$ becomes active is given by the firing rate or gain function $F_i(\vec{n})$, which is an implicit function of the weight function $w_{ij}$ and external inputs $I_i$.  
One of the crucial elements of the ensuing calculation is making the connection between the  gain function $F_i(\vec{n})$, which appears in (\ref{eq:master}) and  $f( I_i(t)+\sum_j w_{ij}n_j )$, which appears in (\ref{eq:wc}).

In general,  we cannot solve for $P(\vec{n},t)$ in (\ref{eq:master}) explicitly. 
One strategy is to derive an expansion  of $P(\vec{n},t)$ in terms of its moments  $\langle n_i(t) n_j(t') n_k(t'') \cdots \rangle$
where the expectation value is over all statistical realizations of the Markov process.  The first moment
\begin{equation}
	a_i(t) = \langle n_i(t) \rangle = \sum_{\vec{n}} n_i P(\vec{n}, t)
\end{equation}
is a measure of the mean activity in the network.  We obtain an equation for $a_i(t)$ by multiplying equation~(\ref{eq:master}) by $n_i$ and taking the sum over all configurations $\vec{n}$.  However,  this equation is not closed (i.e.~it depends upon the second and possibly higher moments).  An equation for the second moment can be similarly constructed by multiplying~(\ref{eq:master}) by $n_i n_j$ and summing over all configurations.  The resulting equation will depend on the third  and higher moments.  Continuing this process will result in a moment hierarchy with as many equations as there are locations, which could be infinite. In general no finite subset of this hierarchy is closed.  This means that if we wish to have a closed set of equations then we need to make some approximation which allows us to truncate the hierarchy.  

The simplest way to close or truncate the moment hierarchy is to assume
that all the higher order moments factorize into products of $a_i(t)$.  This is the naive mean field assumption  where all cumulants are zero.  For example, the second cumulant (i.e. variance) $\langle n_i(t) n_j(t') \rangle-a_i(t) a_j(t')$ is set to zero. For our master equation~(\ref{eq:master}), this assumption yields (see the computation in the next section)
\begin{equation}
	\partial_t a_i(t) = - \alpha a_i(t) +F_i\left (\vec{a} \right) 
	\label{eq:disWc}
\end{equation}
which is similar in form to the rate equation (\ref{eq:wc})  for a discrete domain and with the firing rate function given by $F_i(\vec{a})$.  
However, the fact that statistics in the brain are observed to be near Poisson is devastating for this naive mean field assumption because every cumulant is comparable to the mean, implying any such truncation of the resulting hierarchy is not justifiable.


Here, we describe an alternative means of truncating the hierarchy consistent with near Poisson firing statistics.
 In this case, we observe that the equation for $a_i(t)$ can be written as
 \begin{eqnarray}
	\partial_t a_i(t) &=& - \alpha a_i(t) + F_i\left ( \vec{a} \right) \nonumber \\
	&& + g\left [ \langle n_i(t) n_j(t) \rangle,  \langle n_i(t) n_j(t) n_k(t)\rangle, \langle n_i(t) n_j(t) n_k(t) n_l(t) \rangle, \cdots \right ]
	\label{eq:wcHier1}
\end{eqnarray}
where $g$ is some functional dependent upon the higher moments in the hierarchy.  In order to choose a reasonable approximation and get a finite system of closed equations, we must identify a finite set of higher moments in terms of which we can express the remaining moments.  We are guided by the indication that neuron firing statistics are near Poisson.  Not coincidentally, the solution to the master equation in the case where the gain function $F_i$ is constant or linear is exactly Poisson with mean rate (i.e. stochastic intensity) determined by the Wilson-Cowan rate equation.   In order to truncate the hierarchy, we perform a change of variables to measure the deviations of each cumulant from the value under Poisson statistics.  This new hierarchy is truncatable, as will be demonstrated \emph{a posteriori}.  From the perspective of solving the master equation, this new hierarchy is the natural one because the underlying statistical model is a point process.

The moment hierarchy approach does not make any approximation.  It is a change of variables from the distribution $P(\vec{n},t)$ to moments of that distribution.  The approximation arises when we truncate this hierarchy in order to render the equations tractable.  The simplest truncation is mean field theory.  The first order corrections to mean field theory are given by truncating at the next order.  Truncation of the moment hierarchy requires some justification.  We will demonstrate below that this justification in the neural case may be provided by the large spatial extent of neural connectivity and the distance of the system from a bifurcation.

\section{Truncation of the Moment Hierarchy}
\label{sec:momentHierarchy}

We will derive a moment hierarchy from our master equation~(\ref{eq:master}) and then show how it can be truncated.  To get an equation for the first moment $a_i(t)$, we multiply the master equation~(\ref{eq:master}) by $n_i$ and sum over all configurations $\vec{n}$:
\begin{eqnarray}
	\sum_{\vec{n}} n_i \frac{d P(\vec{n}, t)}{dt} &=& \sum_{\vec{n}} n_i \left \{ \sum_k \alpha (n_k + 1) P(\vec{n}_{k+},t) -  \alpha n_kP(\vec{n},t) \right . \nonumber \\
	&&+ \left . F_i\left (  \vec{n}_{i-} \right) P(\vec{n}_{i-}, t)  - F_i\left (  \vec{n} \right) P(\vec{n}, t) \right \}
	\label{eq:aEq1}
\end{eqnarray}
The first two terms on the right hand side simplify to
\begin{eqnarray}
	\sum_{\vec{n}} n_i \left ( \sum_k \alpha (n_k + 1) P(\vec{n}_{k+},t) -  \alpha n_kP(\vec{n},t) \right )
	&=& \alpha  \sum_{k \ne i} \sum_{\vec{n}}  n_i \left ( n_k P(\vec{n}, t) -  n_k P(\vec{n}, t) \right )  \nonumber \\
	&& +\alpha \sum_{\vec{n}} \left \{ \left ( n_i - 1 \right ) n_i P(\vec{n}, t)  - n^2_i P(\vec{n}, t) \right \} \nonumber \\
	&=& - \alpha \sum_{\vec{n}} n_i P(\vec{n}, t) = - \alpha a_i(t)
\end{eqnarray}
The first equality results from re-indexing the summation over $n_k$ from $(0, \infty)$ to $(1, \infty)$.  We leave the summation indicated as over all configurations $\vec{n}$ because the factor of $n_k$ prevents the $0$ term from contributing.  We have also separated out the terms where $i=k$; the only term which survives is one of these.  Note that we have made no approximations thus far.  The terms involving the function $F_i(\vec{n})$ take the form
\begin{eqnarray}
	\sum_{\vec{n}} n_i \sum_{k}  \left \{ F_i\left (  \vec{n}_{i-} \right) P(\vec{n}_{i-}, t)  - F_i\left (  \vec{n} \right) P(\vec{n}, t) \right \} \nonumber \\
	= \sum_{k\ne i} \left \{ \sum_{\vec{n}} n_i F_k(\vec{n}) P(\vec{n}, t) -  \sum_{\vec{n}} n_i F_k(\vec{n}) P(\vec{n}, t)  \right \} \nonumber \\
	+  \sum_{\vec{n} }  \left ( n_i +1 \right ) F_i(\vec{n}) P(\vec{n}, t)  - \sum_{\vec{n}} n_i F_i(\vec{n}) P(\vec{n}, t) \nonumber \\
	= \sum_{\vec{n}} F_i(\vec{n}) P(\vec{n}, t)
\end{eqnarray}
  Unlike the first term, we cannot directly represent this term as a function only of $a_i(t)$, due to its nonlinear nature.  The equation for the mean is therefore:
  \begin{eqnarray}
  	\frac{d a_i}{dt} = -\alpha a_i + \langle F_i\left (\vec{n} \right ) \rangle  \label{eq:N1}
  \end{eqnarray}
where $\langle F_i(\vec{n})\rangle$ will  include higher order moments.
  
We continue by constructing an equation for the second moment 
\begin{equation}
	N_{ij} = \langle n_i n_j \rangle = \sum_{\vec{n}} n_i n_j P(\vec{n}, t) 
\end{equation}
and third moment
\begin{equation}
	N_{ijk} = \langle n_i n_j n_k\rangle = \sum_{\vec{n}} n_i n_j n_k P(\vec{n}, t) 
\end{equation}
to obtain
\begin{eqnarray}
	\frac{d N_{ij}}{dt} &=&  - 2\alpha N_{ij} + \alpha \delta_{ij} a_i + \langle n_j F_i + n_i F_j \rangle + \delta_{ij} \langle F_i \rangle \label{eq:N2}\\
	\frac{d N_{ijk}}{dt} &=& -3\alpha N_{ijk} + \alpha \langle n_i n_j \delta_{im} + n_i n_m \delta_{jm} + n_j n_m \delta_{ij} \rangle - \alpha \delta_{ij} \delta_{jm} \langle n_i \rangle \label{eq:N3} \\
	&+& \langle n_i n_j F_k + n_i n_k F_j + n_j n_k F_i \rangle + \langle n_j F_k \delta_{ik} + n_k F_i \delta_{ij} + n_i F_j \delta_{kj} \rangle\\ & -& 3\langle n_i^2 F_i \rangle \delta_{ij} \delta_{jm} \nonumber
\end{eqnarray}

Since we expect solutions to be near Poisson, we transform the hierarchy to describe the departure of moments from Poisson statistics.  For a Poisson distribution, the cumulants are all equal to the mean of the distribution.   Hence, we introduce what are called ``normal ordered cumulants", which measure the ``deviations" of the cumulants from Poisson values.  The first normal ordered cumulant is the same as the first moment $a_i(t)$.  The next two are given by
\begin{equation}
	C_{ij} = N_{ij} -a_i a_j  - a_i \delta_{ij} 
	\label{eq:2ndnoc}
\end{equation}
and
\begin{eqnarray}
	C_{ijk} &=& N_{ijk}  - N_{ij} a_k - N_{jk} a_i - N_{ik} a_j +2 a_i a_j a_k  \nonumber  \\
	&&- ( N_{ij} - a_i a_j ) \delta_{jk} - ( N_{ik} - a_i a_k) \delta_{ij} - (N_{jk} - a_j a_k ) \delta_{ik} + 2a_i \delta_{ij} \delta_{jk}
	\label{eq:3rdnoc}
\end{eqnarray} 
The normal ordered cumulants can be computed using a recursive algorithm.  The algorithm involves replacing all moments $N_{ijk\cdots}$ with ``normal-order-corrected'' moments recursively by subtracting terms with coincident or ``contracted" indices which reduce the order of the moment.  For example, the ordinary second cumulant is simply $N_{ij}-a_i a_j$.  To compute the normal ordered version, we replace the moments appearing in the expression with the normal-ordered-corrected forms, i.e. set $N_{ij}\rightarrow N_{ij}-a_i\delta_{ij}$.  The term subtracted results from the contraction of the $i,j$ indices, i.e. $N_{ij} \rightarrow N_i \delta{ij} = a_i \delta_{ij}$.  The third cumulant is more complicated but follows the same strategy.  The important thing to note here is that the higher moments must be independently corrected for each group of contracted indices.  The ordinary third cumulant is given by
\begin{equation}
	\langle \left ( n_i - a_i \right )\left ( n_j - a_j \right )\left ( n_k - a_k \right )\rangle =  N_{ijk}  - N_{ij} a_k - N_{jk} a_i - N_{ik} a_j +2 a_i a_j a_k
\end{equation}
To obtain the normal ordered cumulant, we first make the replacement $N_{ijk}\rightarrow N_{ijk}-N_{ij}\delta_{jk}-N_{ik}\delta_{ij}-N_{jk}\delta_{ik}  - a_i\delta_{ij}\delta_{jk}$.  We then must correct for all appearances of the second moment $N_{lm}$ resulting in (\ref{eq:3rdnoc}).  This algorithm systematically removes the underlying Poisson contributions (at all tensor ranks) and leaves us with the normal ordered cumulants.  For a Poisson distribution, all $C_{ijk \cdots} = 0$ except the first, $a_i$.   An alternative rationale for normal ordering will appear in Sec. \ref{sec:field}.
 

Transforming the first three equations of the hierarchy (\ref{eq:N1}), (\ref{eq:N2}), and (\ref{eq:N3})  yields
  \begin{eqnarray}
  	\frac{d a_i}{dt} &=& -\alpha a_i + \langle F_i\rangle  \label{eq:a}\\
	\frac{d C_{ij}}{dt} &=&  - 2\alpha C_{ij} + \langle (n_j-a_j) F_i\rangle + \langle (n_i-a_i) F_j \rangle  \\
	\frac{d C_{ijk}}{dt} &=&  - 3\alpha C_{ijk} + \langle (n_i - a_i)(n_j-a_j) F_k + {\rm permutations} \rangle
  \end{eqnarray}
where we must re-express the expectation values involving $F_i$ in terms of the normal ordered cumulants.  

Since $F_i(\vec{n})$ is defined in terms of the vector of active neuron numbers, $\vec{n}$, its expectation value will be naturally expressed in terms of the moments, $N_{ijk\cdots}$, as given by the Taylor expansion of $F(\vec{n})$.
\begin{equation}
	\langle F_i(\vec{n}) \rangle= \sum_j F_i^j a_j + \frac{1}{2} \sum_{jk} F_i^{jk} N_{jk} + \cdots
\end{equation}
We have implicitly defined the notation that $F^{jk\cdots}_i$ is the derivative of $F_i$ with respect to $n_j, n_k, \cdots$.  Note that this expansion only applies to the expectation value of $F_i(\vec{n})$.  We need to re-express this series  as an expansion  in terms of the normal ordered cumulants.  This transformation of variables will rearrange the terms and result in a new series with new coefficients that sums to the same result as the original series.
For example every term proportional to $N_{ijk\cdots}$ will contribute a term proportional to $a_i a_j a_k \cdots$ to the normal ordered cumulant expansion.  This means that we can write the expansion in the form
\begin{equation}
	\langle F_i \rangle = F_i(\vec{a}) + \cdots
\end{equation}
as we would expect.  However, there are also contributions from the normal ordering corrections.  The simplest are those which arise due to every index being coincident or contracted at each order.  This produces a term linear in $a_i$.  These corrections form the series
\begin{equation}
	\langle F_i \rangle = F_i(\vec{a}) + \sum_{m=2} \sum_j \frac{F^{j^m}_i}{m!} a_j  + \cdots
	\label{eq:noseries}
\end{equation}
where by $F^{j^m}$ we mean the $m$th derivative of $F(\vec{n})$ with respect to $n_j$.  Were $F_i(\vec{n})$ a polynomial this procedure would truncate at the highest order of the function. However, for an arbitrary general function these corrections quickly become unwieldy as one proceeds through the orders of the expansion to include corrections to the terms that go as $a_ia_j$, $a_i a_j a_k \cdots$.

Our perspective has been to interpret the Wilson-Cowan equation as describing the mean field of some Poisson process with activating and decaying transitions.  Hence, the Wilson-Cowan equation should be the mean field solution of the normal-ordered cumulant hierarchy (\ref{eq:a}).  However, the gain function in the Markov equation is not the same as the gain function in the Wilson-Cowan equation.  The Wilson-Cowan gain function is the normal ordered version of the Markov gain function.  Thus, in order for the Wilson-Cowan gain function to have the form of $f(s_i)$, where $s_i = \sum_{i} w_{ij} a_j + I_i(t)$ we 
 assume that this re-summation produces 
 \begin{equation}
	\langle F_i \rangle =  f(s_i)+ h.o.t.
\end{equation} 
where 
and the higher order terms ($h.o.t.$) are dependent upon the higher normal ordered cumulants according to the Taylor series expansion of $f(s)$.  It will be seen in Sec.~\ref{sec:field} that there always exists a Master equation gain function $F$ such that $f$ is expressible in this form and the resummation works
to produce the same $f(s)$ (and derivatives thereof) at every order in this expansion.

We now return to the series (\ref{eq:noseries}) to consider terms with precisely one factor of $C_{ij}(t)$.  At each order $m$ in the series (i.e. the order which contains the $m$th moment), there are $m(m-1)/2$ terms which have one factor of $C_{ij}(t)$.  These terms sum to give
   \begin{eqnarray}
  	\sum_{m=2} \frac{f^m(I_i(t))}{m!} \frac{m(m-1)}{2} \left ( \sum_j w_{ij} a_j(t) \right )^{m-2} \sum_{kl}w_{ik} w_{il} C_{kl} 
	\label{eq:seriesC}
  \end{eqnarray}
which can be rewritten as
\begin{eqnarray}
\frac{1}{2} f''\left ( \sum_{j} w_{ij} a_j(t) + I_i(t) \right ) \sum_{jk} w_{ij} w_{ik} C_{jk}(t)
\end{eqnarray}
Other terms are at least second order in $C_{ij}$, higher normal ordered cumulants, or corrections from normal ordering.
Our first generalized activity equation excluding terms dependent on third and higher normal ordered cumulants is therefore
\begin{eqnarray}
	\partial_t a_i(t) &=& - \alpha a_i(t) + f \left ( \sum_j w_{ij} a_j(t) + I_i(t) \right) \nonumber \\
	&&+ \frac{1}{2} f''\left ( \sum_{j} w_{ij} a_j(t) + I_i(t) \right ) \sum_{jk} w_{ij} w_{ik} C_{jk}(t)
	\label{eq:genDisWc}
\end{eqnarray}

We can take this same approach in order to compute an equation for $C_{ij}(t)$ and obtain
\begin{eqnarray}
	\frac{d}{dt}C_{ij}(t) &=& -2\alpha C_{ij}(t) + f' \left ( s_i \right ) \sum_{k} w_{ik}C_{kj}(t) + f'\left ( s_j \right ) \sum_{k} w_{jk} C_{ki}(t) \nonumber \\
	&&+ f' \left ( s_i \right ) w_{ij} a_j(t) + f' \left ( s_j \right ) w_{ji} a_i(t) 
	\label{eq:genWcC}
\end{eqnarray}
Equations~(\ref{eq:genDisWc}) and (\ref{eq:genWcC}) constitute a closed set of equations for the mean and variance of a Wilson-Cowan network of neurons.  $a_i(t)$ represents the Poisson rate and $C_{ij}(t)$ measures the deviation of the variance from Poisson statistics.
These equations are the minimal consistent extension of the Wilson-Cowan rate equations to include the effects of higher order statistics.   Higher order corrections can be incorporated by adding terms involving higher order cumulants into (\ref{eq:genDisWc}) and  (\ref{eq:genWcC}) and including equations for these higher order cumulants.    We also note that the gain function need not be analytic everywhere for this expansion to work.  It can contain a countable number of non-continuous or non-differentiable points.  The equation would be corrected with the inclusion of impulse function terms at these singular points.

An immediate noteworthy consequence is that $C_{ij}(t)$ will only have substantial input when the activity is such that $f'(s)$ is large.   As an example suppose $f(s)$ is a simple sigmoid function.  In this case, $f'(s)$ is peaked at threshold (where we define threshold to be the central point of half maximum) and zero far away from threshold.  Reasonably, we have the result that correlated activity will only increase when the input to a neuron is near threshold. If the slope of the sigmoid  is such that $f(s)$ is a step function, or near to a step function, then $C_{ij}(t)$ will receive input only when the activity is precisely near threshold.  Also notice that the strength of the input to $C_{ij}(t)$ is proportional to the weight $w_{ij}$ between the neurons in question as well as the mean activity.  An initial check on the equations is that $C_{ij}$ decouples from $a_i$ in the case where $f(s)$ is linear or constant.

To consider the dynamics of large scale neural activity, we can take the continuum limit of these equations to get equations for the mean activity $a(x,t)$ and correlation $C(x,y,t)$.
\begin{eqnarray}
	\partial_t a(x,t) &=& - \alpha a(x,t) + f \left ( \int dy w(x,y) a(y,t) + I(x,t) \right) \nonumber \\
	&&+ \frac{1}{2} f''\left ( \int dy w(x,y) a(y,t) + I(y,t) \right )\nonumber \\
	&& \times   \int dy dz w(x,y) w(x,z) C(y,z,t)
	\label{eq:genConWc}
\end{eqnarray}
and 
\begin{eqnarray}
	\frac{d}{dt}C(x,y,t) &=& -2\alpha C(x,y,t) + f' \left ( s(x) \right ) \int dz w(x,z)C(z,y,t) \nonumber \\
	&&+ f'\left ( s(y) \right ) \int dz w(y,z) C(z,x,t) \nonumber \\
	&&+   f' \left ( s(x) \right ) w(x,y) a(y,t) + f' \left ( s(y) \right ) w(y,x) a(x,t) 
	\label{eq:genConWcC}
\end{eqnarray}
These are the generalized activity equations.  Had we wished to include even higher moments we could have continued through the hierarchy.  For simplicity of illustration, we truncate the hierarchy at this level.

\subsection{Criticality and truncation of the hierarchy}
\label{sec:crit}

Although we have derived equations for the mean activity and equal-time correlation, there are some outstanding issues.  The primary concern which must be addressed is that we require some justification for the truncation of the hierarchy at the level of the two-point correlation function $C_{ij}(t)$ instead of allowing higher moments to interact with the mean activity.

Consider the mean field equation without the correction due to correlated activity~(\ref{eq:disWc}).
Define $a^0_i$ to be some steady state solution to this equation and linearize equation~(\ref{eq:disWc}) around this solution.  The  perturbations $\delta a_i(t)$, from this steady state solution obey the equation
\begin{equation}
	\partial_t \delta a_i(t) = - \alpha \delta a_i(t) + f' \left ( \sum_j w_{ij} a^0_j + I_i(t) \right) \sum_{j} w_{ij} \delta a_j(t) 
	\label{eq:linWc}
\end{equation}
We rewrite this equation as
\begin{equation}
	\partial_t \delta a_i(t) = - \sum_j \Gamma_{ij}[a^0] \delta a_j(t) 
\end{equation}
where the matrix $\Gamma_{ij}$ is defined by
\begin{equation}
	\Gamma_{ij}[a^0] =  \alpha \delta_{ij} - f' \left ( \sum_j w_{ij} a^0_j + I_i(t) \right) w_{ij}
\end{equation}
If all of the eigenvalues of $\Gamma_{ij}$ are positive, then the solution $a^0_i$ is stable.  Likewise, negative eigenvalues indicate instability.  Criticality is the condition of marginal stability, in which one or more of the eigenvalues are $0$.

Returning to the equation for $C_{ij}(t)$, we see
\begin{eqnarray}
	\frac{d}{dt}C_{ij}(t) &=& -\sum_k \left ( \Gamma_{ik} [a(t)]C_{kj}(t) + \Gamma_{jk} [a(t)]C_{ik}(t) \right )  \nonumber \\
	&&+  f' \left ( s_i \right ) w_{ij} a_j(t) + f' \left ( s_j \right ) w_{ji} a_i(t) 
\end{eqnarray}
We assume that the mean field solution $a^0_i$ is stable.  In addition, we assume, per the truncation hypothesis, that the steady state value of $C_{ij}(t)$ does not appreciably alter either $a_i$ or, therefore, the matrix $\Gamma_{ij}$.  In this case, $\Gamma_{ij}$ has all positive eigenvalues and is diagonalizable.   Define
\begin{equation}
	\Lambda_{ij} = \lambda_i \delta_{ij} = \sum_{lk} U_{il} \Gamma_{lk} U^{-1}_{kj}
\end{equation}
to be the diagonalization of $\Gamma_{ij}$.  We also define the shorthand
\begin{equation}
	A_{ij} = f' \left ( s_i \right ) w_{ij} a_j(t) + f' \left ( s_j \right ) w_{ji} a_i(t) 
\end{equation}
for the driving terms in equation~(\ref{eq:genWcC}). 
The steady state solution is given by
\begin{equation}
	C^0_{ij} = \sum_{\bar{l}l \bar{k}k } U^{-1}_{i\bar{k}} U^{-1}_{j\bar{l}} \left ( \lambda_{\bar{l}}+ \lambda_{\bar{k}}\right )^{-1} U_{\bar{l}l} U_{\bar{k}k} A_{kl}
	\label{eq:disCsoln}
\end{equation}
Notice that each term contributing to the magnitude of $C^0_{ij}$ is attenuated by a sum of eigenvalues.  The magnitude of the eigenvalues $\lambda_i$ determines the distance of the system from a bifurcation or criticality, i.e. it is the distance from the onset of an instability.  Thus, the further the system is from criticality the more attenuated the fluctuations and the more justified we are in truncating the hierarchy.  Conversely, the closer the system is to criticality the more the approximation breaks down.  At criticality, this solution~(\ref{eq:disCsoln}) becomes singular.  This is an indication that criticality is a fluctuation dominated, as opposed to mean field dominated, regime.  A similar argument will extend to any equation in the hierarchy and we are left with an intuitively satisfying result:  stability smooths out fluctuations.  This argument is what allows us to disregard the effects of still higher moments upon the mean activity and truncate by considering only the two-point correlation function's effect on the mean.

It is also worth noting that the eigenvalues relevant for the dynamics of the two point correlation $C_{ij}$ are the sums of the eigenvalues of the mean field equation, $\lambda_i + \lambda_j$.  In the case that $a^0_i$ is stable, not only will $C_{ij}$ be stable but it will relax to equilibrium faster in general than the mean field solution.  In kinetic theory, this is akin to the Bogoliubov approximation, in which the collision term is computed by solving for the steady state of the two point correlation on the assumption that the correlation function reaches steady state on a time scale shorter than the mean field.  In our case, this approximation leads to
\begin{eqnarray}
	\partial_t a_i(t) &=& - \alpha a_i(t) + f \left ( \sum_j w_{ij} a_j(t) + I_i(t) \right) \nonumber \\
	&&+ \frac{1}{2} f''\left ( \sum_{j} w_{ij} a_j(t) + I_i(t) \right ) \sum_{jk} w_{ij} w_{ik} C^0_{jk}
	\label{eq:genDisWcCollision}
\end{eqnarray}
Note immediately that the input from correlated activity decouples in the case that the firing rate function is in a linear or constant region, since the coupling is proportional to the second derivative of $f(s)$.  

The extent of neural interconnections also has an effect on the size of correlations and the ability to truncate the hierarchy.  Consider that neuron $i$ has $N_i$ pre-synaptic neurons (i.e. neurons for which $w_{ij} \ne 0$).  Further, let the average connectivity weight over all inputs be $w_0$.  In this case we can approximate $w_{ij} \approx w_0/N_i$.  The steady state value of $C_{ij}(t)$, which is determined by the driving term $A_{ij}(t)$, is seen to be inversely proportional to the number of pre-synaptic neurons due to the linear dependence on $w_{ij}$ of $A_{ij}$.  In the most extreme case the number of pre-synaptic neurons is the entire network, so $N_i = N$, and the correlation function $C_{ij}(t)$ becomes simply a finite size effect, going  as $1/N$.  Smaller system sizes in general will have larger correlations, which is intuitively sound.  More generally, as long as we can bound the total input to any given neuron, we can define $N_m = {\rm min}N_i$ and scale all weights so that they can be written
\begin{equation}
	 w_{ij} \approx \frac{N_m}{N_i} \frac{w_M}{N_m}
\end{equation}
where $w_M$ is the maximum total input to any given neuron.
This allows us to treat $N_m$ as an effective system size.  Larger $N_m$ reduces the effects of fluctuations at a given distance to a bifurcation.


We have two competing effects.  On one hand, we have the system size governing the magnitude of correlations.  On the other, the distance of the system to a bifurcation likewise affects the size of fluctuations.  The relative tradeoff of the two, from the definition of $C^0_{ij}$ is given by the product of the smallest eigenvalue of $\Gamma_{ij}$ and the number of pre-synaptic neurons  $N_i$.
We will demonstrate this relationship more precisely when considering the all-to-all network in section~\ref{sec:alltoall}.

Finally, it is worth pointing out the effect of input upon the hierarchy.  If the input is another Poisson process then the only equation which is affected is the equation for $a_i(t)$.  The higher equations in the hierarchy are only affected by this input through its effect on the mean activity $a_i(t)$ and the firing rate function $f(s)$.  In general, this suggests that large external inputs will actually reduce fluctuations, depending on the form of $f(s)$, in the sense of driving the system towards Poisson-like behavior, a reasonable result.  In particular, if $f(s)$ is a saturating function, then the correlations will decouple from the equation for $a_i(t)$ and the source terms for higher correlations will be driven to zero, leaving the system described completely by the rate equations.  The analogous situation for a ferromagnet is driving the system with a large external magnetic field.




\section{Path Integral Solution of the Master Equation}
\label{sec:field}


We have thus far demonstrated how a minimal Markov model consistent with the Wilson-Cowan rate equation can be used to derive generalized equations in a hierarchy of moments. Although we truncated this hierarchy at second order, one can in principle truncate at any desired cumulant, although the calculations become successively more cumbersome.  Here we show that the moment hierarchy is equivalent to a path integral or field theoretic approach, which systematizes the perturbation theory for the statistics of the network by providing rules for the construction and evaluation of the cumulants.  Another major benefit is that it provides a systematic means for obtaining moment truncations or closures.  The path integral representation of the master equation~(\ref{eq:master}) was derived by Buice and Cowan~\cite{bc} by modifying a method originally developed for reaction diffusion systems~\cite{doi1,doi2,peliti}.  We quickly review the representation and then detail how the generalized equations can be derived from this representation.  

The moment generating function for the probability density $P(\vec{n},t)$ is given by
\begin{equation}
Z[J_i]=\sum_{\vec{n}} P(\vec{n},t) \exp\left(\sum_i J_i n_i\right)
\end{equation}
where the sum is over all configurations of $\vec{n}$.
Moments of $P$ are obtained by taking derivatives of the generating function with respect to $J_i$.  For example $\langle n_in_j\rangle = \partial^2 Z/\partial J_i\partial J_j |_{\vec{J} = 0}$.  The cumulants can be obtained by taking the derivatives of $W[J]=\ln Z$.  
Field theory generalizes the generating function over a set of discrete variables to a generating functional over functions or fields.  The result is a functional or path integral over all possible paths of time evolution for the system, weighted by the probability of that particular evolution. While it is sometimes possible and desirable to take the spatial continuum limit of the neural system, this is not necessary for the path integral approach.  Here we use continuum spatial notation, although the results carry through in the case where $x$ indexes a discrete variable.   Buice and Cowan~\cite{bc} showed that the generating functional for the master equation~(\ref{eq:master}) is given by 
\begin{equation}
	Z[J(x,t), \tilde{J}(x,t)] \equiv e^{W[J(x,t), \tilde{J}(x,t)]}=\int {\cal D}\tilde{\varphi} {\cal D} \varphi e^{-S[\tilde{\varphi}, \varphi]}e^{\tilde{\varphi}\cdot J + \varphi \cdot \tilde{J}}
	\label{eq:genfun}
\end{equation}
where 
\begin{equation}
	S[\tilde{\varphi}, \varphi] = \int d^dx dt \left [ \tilde{\varphi} \frac{\partial}{\partial t} \varphi  + \alpha \tilde{\varphi} \varphi  - \tilde{\varphi} f\left ( w * \left \{ \tilde{\varphi} \varphi + \varphi \right \} \right) \right ] -  W[\tilde{\varphi}(x,0)]
	\label{eq:action}
\end{equation}
is called the action and we use the notation $u\cdot v = \int d^dx dt\, u(x,t) v(x,t)$.  The fields $\varphi$ and $\tilde{\varphi}$, which are obtained from the configuration variables $\vec{n}$, are defined below  \cite{Janssen2005147,0305-4470-38-17-R01}.  
The asterisk denotes convolution of the weight function with the inputs and the term $W[\tilde{\varphi}(x,0)]$ is the cumulant generating functional of the initial distribution and  takes into account arbitrary distributions in the initial condition.  For example, if the initial state is described by Poisson statistics, we have $W[\tilde{\varphi}(x,0)] = \int d^dx\, a_0(x) \tilde{\varphi}(x,0)$, where $a_0(x)$ is the mean of the Poisson distribution at $x$. 
   Analogous to the generating function for discrete variables, functional derivatives with respect to  $\tilde{J}(x,t)$ yield the normal ordered cumulants, such as
\begin{equation}
	\left . \frac{\delta}{\delta \tilde{J}(x,t)}W[J(x,t), \tilde{J}(x,t)] \right |_{J, \tilde{J} = 0} = \langle \varphi(x,t)  \rangle = a(x,t)
\end{equation}
and
\begin{equation}
	\left . \frac{\delta}{\delta \tilde{J}(x,t)}\frac{\delta}{\delta \tilde{J}(x',t')}W[J(x,t), \tilde{J}(x,t)] \right |_{J, \tilde{J} = 0} = \langle \varphi(x,t) \varphi(x',t') \rangle -\langle \varphi(x,t)\rangle \langle \varphi(x',t') \rangle= C(x,t;x',t')
\end{equation}
Within this formalism~\cite{doi1,doi2,peliti,bc},  expectation values of products of  $\varphi$ are the normal ordered cumulants found in the moment hierarchy. 
The normal ordered cumulant $C(x,y,t)$ from (\ref{eq:2ndnoc}) results from setting $t=t'$ in $C(x,t;x',t')$.
The field $\tilde{\varphi}(x,t)$ is a ``response" field and expectation over functions of it yield Green's functions or response functions for the dynamics~\cite{martin}.   The Ito convention is taken for the Langevin equation (\ref{eq:langevin}) so that moments that involve combinations of $\tilde{\varphi}$ and $\varphi$ that are evaluated at the same time are zero.  More specifically, the convention taken is such that $\langle \tilde{\varphi}(x,t')\varphi(x,t)\rangle=0$ if $t\le t'$.

%
We can heuristically derive the action~(\ref{eq:action}), which was derived explicitly in \cite{bc}, and show that it represents a minimal model of the Wilson-Cowan equation where the activity is to be interpreted as a stochastic intensity or rate of a Poisson process.  
Consider an effective Wilson-Cowan Langevin equation
\begin{equation}
\partial_t n=-\alpha n + F(n) + n_0(x)\delta(t-t_0) + \xi(x,t)
\label{eq:langevin}
\end{equation}
where $n(x,t)$ is the neural activity at location $x$ and time $t$, $\xi(x,t)$ is an effective stochastic forcing with probability density functional $P[\xi]$ and the firing rate function $F$ is not necessarily that which appears in equation~(\ref{eq:wc}), but rather is that in the master equation (\ref{eq:master}). We will show that Poisson noise is necessary to match the Buice and Cowan action (\ref{eq:action}).  The probability density functional for $n(x,t)$ can be written formally as
\begin{equation}
P[n]\propto \int {\cal D}\xi \delta\left[\frac{\partial}{\partial t} n  + \alpha n  -  F\left (   n  \right) -n_0(x)\delta(t)  - \xi\right]P[\xi ]
\label{eq:Pphi}
\end{equation}
where $\delta[\cdot]$ is the functional generalization of the Dirac delta function.  The probability density (\ref{eq:Pphi}) constrains the field $n(x,t)$ to obey the Langevin equation~(\ref{eq:langevin}) with initial condition $n_0(x)$.  The delta functional is defined by the generalized Fourier transform
\begin{equation}
P[n] \propto \int {\cal D} \xi {\cal D}  \tilde{n}  \exp\left({-\int d^dxdt\tilde{n}(\partial_t n+\alpha n-F( n) - n_0 )-\int d^dxdt\, \tilde{n}\xi }\right)P[\xi]
\end{equation}
and $\tilde{n}$ is integrated along the imaginary axis.  
We can now integrate over the stochastic variable $\xi$ to obtain a noise generating functional defined by
\begin{equation}
e^{W[\tilde{n}]}=\int {\cal D}\xi \exp\left({-\int d^dxdt \tilde{n}\xi }\right) P[\xi]
\end{equation}
We choose $\xi$ such that 
\begin{equation}
W[\tilde{n}]=(e^{\tilde{n}} -1-\tilde{n})F+\alpha n(e^{-\tilde{n}} -1+\tilde{n})
\end{equation}
which is consistent with a Poisson activation at rate $F$ and a Poisson decay at rate $\alpha$.
We next  transform to the new variables
\begin{eqnarray}
	\varphi(x,t) &=& n(x,t) e^{-\tilde{n}(x,t)} \nonumber \\
	\tilde{\varphi}(x,t) &=& e^{\tilde{n}(x,t)} - 1
	\label{eq:ntophi}
\end{eqnarray}
The transformation (\ref{eq:ntophi}) to the new fields serves to simplify the noise generating functional and results in an action that has the form of (\ref{eq:action}) but with a different gain  function.  This new action is reconciled with (\ref{eq:action}) by the normal ordering operation.  The reason this is necessary is because the Ito convention used to interpret the action would not hold uniformly for the transformed fields since in performing the transformation (\ref{eq:ntophi}), the $\tilde{\varphi}$ and $\varphi$ fields  inside the gain function are evaluated at the same time and moments between these particular instances of the fields would not necessarily be zero.  This inconsistency can be corrected by redefining (i.e. normal ordering) the terms in the gain function.
%
As an example, consider the firing rate function to be $F(n) = (w\cdot n)^2$.  After transforming, it becomes $F = (w \cdot (\tilde{\varphi} \varphi + \varphi) )^2$, which
mixes response and configuration operators at the same time point.  To restore Ito convention, we normal order so that response and configuration variables are no longer mixed.  We do this by considering the $n(x,t)$ operators at separate times $t, t'$ and computing how the operators approach each other as $t \rightarrow t'$. The properties of the response field provide $\lim_{t \rightarrow t'+}\langle \varphi(x,t) \tilde{\varphi}(x',t') \rangle = \delta(x - x')$ and $\lim_{t \rightarrow t'-}\langle \varphi(x,t) \tilde{\varphi}(x',t') \rangle =0$ so that $F$ becomes 
\begin{eqnarray}
	\lim_{t \rightarrow t'}\langle(w \cdot (\tilde{\varphi}(x,t) \varphi(x,t) + \varphi)(x,t) )(w \cdot (\tilde{\varphi}(x,t') \varphi(x,t') + \varphi)(x,t') )\rangle = \nonumber \\
	 \langle(w \cdot (\tilde{\varphi}(x,t) \varphi(x,t) + \varphi)(x,t) )^2 + w^2 \cdot (\tilde{\varphi}(x,t) \varphi(x,t) + \varphi)(x,t) )\rangle
\end{eqnarray}
Hence, restoring the Ito convention requires the replacement $n^2 \rightarrow n^2 + n$
(albeit in the variables $\varphi, \tilde{\varphi}$ and similarly for higher powers of $n$).  This normal ordering will adjust the gain function to be $f(w*(\tilde{\varphi}\varphi+\varphi))$, leaving us with the action~(\ref{eq:action}).  From the perspective of the original master equation~(\ref{eq:master}), the transformation to $\varphi, \tilde{\varphi}$ is equivalent to expanding solutions to the master equation around a Poisson solution.

\subsection{Closed activity equations from the path integral}

We derive the generalized activity equations for the normal ordered cumulants directly from the generating functional by Legendre transforming to an effective action and then calculating the extrema of the effective action~\cite{zinnjustin,cornwall,bc}.  
We first perform the computation for the mean activity $a(x,t)$ and then show how to generalize to arbitrary numbers of cumulants.
	The generating functional (\ref{eq:genfun}) can be written more compactly as
\begin{equation}
		\exp\left(W[ J^{\mu}(x,t) ]\right)= \int {\cal D}\Phi_{\mu} \exp{\left (-S[\Phi_{\mu}] + \int d^dx \,dt\, J^{\mu}(x,t) \Phi_{\mu}(x,t)  \right )}  
\label{eq:gf}
\end{equation}
where we define $\Phi_{\mu}(x,t)$, where $\mu\in \{1, -1\} $ such that  $\Phi_{1}(x,t) =\varphi(x,t)$ and $\Phi_{-1}(x,t) = \tilde{\varphi}(x,t)$.  We use Einstein summation convention (i.e. when the same index appears twice (one upper, one lower) in equations, summation will be implied).  Similarly we define $J_{\mu}(x,t)$ via $J_{1} = J$ and $J_{-1} = \tilde{J}$.  We can ``raise" an index $\mu$ via multiplication with the matrix 
\begin{equation}
	J^{\mu} = \left ( \begin{array}{cc} 0 & 1 \\ 1 & 0 \end{array} \right) J_{\mu}
\end{equation}
so that $J^1 = \tilde{J}$ and $J^{-1} = J$.  
	Thus we have $J^\mu \Phi_\mu = J^1 \Phi_1 + J^{-1} \Phi_{-1} = \tilde{J}\varphi + J \tilde{\varphi}$.  
In order to streamline our notation, we will also define the dot product as the integral
\begin{equation}
	J^{\mu} \cdot a_\mu =  \int d^dx dt\, J^\mu(x,t) a_\mu(x,t) 
\end{equation}
For functions of more than one spatial variable, this inner product notation will generalize to the trace of the matrix product, i.e.
\begin{equation}
	A(x,y,t) \cdot B(x,y,t) = \int dt dx dy A(x,y,t) B(x,y,t)
\end{equation}
We will also write the action as $S[\Phi_\mu]=\int d^dx\, dt\, L[\Phi_\mu]$, where $L$ is the integrand of the action in (\ref{eq:action}).

Define $a_\mu = \langle \Phi_{\mu} \rangle$. The ``effective action" $\Gamma[a_\mu]$ for $a_\mu$ is derived by a Legendre transformation
\begin{equation}
\Gamma[a_\mu(x,t)]=  J^\mu(x,t) \cdot a_\mu(x,t) - W[J^\mu(x,t)]
\label{eq:legendre}
\end{equation}
where the conditions
\begin{eqnarray}
\frac{\delta W[J^\mu(x',t')]}{\delta J^\mu(x,t)}&=&a_\mu(x,t)\\
\frac{\delta \Gamma[a_\mu(x',t')]}{\delta a_\mu(x,t)}&=&J^\mu(x,t) \label{eq:J}
\end{eqnarray}
are enforced.  In analogy with  classical mechanics, the extrema of the effective action 
\begin{equation}
\frac{\delta \Gamma[a_\mu(x',t')]}{\delta a_\mu(x,t)}=0
\end{equation}
give the equations of motion or the activity equations for $a_\mu(x,t)$.  

In general, we will not be able to compute the equation of motion exactly since the path integral in (\ref{eq:gf}) cannot be computed exactly.  However, we can perform a steepest descent asymptotic expansion of (\ref{eq:gf}) and compute the activity equation perturbatively.   In field theory, this  is known as the loop expansion because the terms in the expansion can be represented by Feynman diagrams whose order is given by the number of loops that diagram possesses.  Substituting for $W[J^\mu]$ using the Legendre transformation (\ref{eq:legendre}) gives
\begin{equation}
\exp(-\Gamma[a_\mu])= \int {\cal D}\Phi_{\mu} \exp{\left (-S[\Phi_{\mu}] + J^{\mu} \cdot \Phi_{\mu} -J^\mu \cdot a_\mu\right )}
\end{equation}
where we have suppressed the $x$ and $t$ arguments.  Defining a new variable $\Psi_\mu=\Phi_\mu-a_\mu$ and using (\ref{eq:J}) gives
\begin{equation}
\exp(-\Gamma[a_\mu])= \int {\cal D}\Psi_{\mu} \exp\left (-S[\Psi_{\mu}+a_\mu] +  \Gamma^{\mu} \cdot\Psi_{\mu} \right )
\end{equation}
where we define $\Gamma^\mu[a_\mu] \equiv \delta\Gamma/\delta a_\mu$.  We now expand $S[\Psi_\mu+a_\mu]$  in a functional Taylor series to obtain
\begin{equation}
S[\Psi_\mu+a_\mu]= S[a_\mu]+L^{\mu}[a_\mu]\cdot \Psi_\mu + \frac{1}{2} L^{\mu \nu}[a_\mu]\cdot\Psi_\mu \Psi_\nu+ V[\Psi_\mu, a_\mu]
\end{equation}
where $L^{\mu}$ represents the functional derivative of $L[a_\mu]$ with respect to $a_{\mu}$ and similarly for $L^{\mu \nu}$ and higher derivatives. $V[\Psi_\mu, a_\mu]$ represents the remaining terms in the Taylor expansion.  By definition, these are at least cubic in $\Psi_\mu$. 
Hence
\begin{equation}
\exp(-\Gamma)= \int {\cal D}\Psi_{\mu} \exp{\left (-S[a_\mu]-L^{\mu}[a_\mu]\cdot\Psi_\mu- \frac{1}{2} L^{\mu \nu}[a_\mu]\cdot\Psi_\mu\Psi_\nu-V +  \Gamma^{\mu}\cdot \Psi_{\mu}  \right )}
\end{equation}
We introduce a scaling parameter for the action, $h$, (which we will set equal to one) via
\begin{eqnarray}
	S &\rightarrow& \frac{1}{h} S \nonumber \\
	\Gamma &\rightarrow& \frac{1}{h} \Gamma
\end{eqnarray}
 We will show that an expansion in powers of $h$ is consistent with the truncation used in deriving the moment hierarchy in section~\ref{sec:momentHierarchy}.  The reason is that it organizes the terms in the expansion so that the true small parameters in the system, namely inverse distance from criticality and inverse number of inputs, are manifested.  We thus consider an asymptotic expansion  $\Gamma= \Gamma_0 +h\Gamma_1 +h^2\Gamma_2+\cdots$. 
If we set  $\Gamma_0=S$ we obtain
\begin{equation}
\exp \left (- \frac{1}{h}\Gamma \right )= \exp \left (- \frac{1}{h}S[a_\mu]\right )\int {\cal D}\Psi_{\mu} \exp{\left (- \frac{1}{2h} L^{\mu \nu}[a_\mu]\cdot\Psi_\mu\Psi_\nu - \frac{1}{h}V[\Psi_\mu, a_\mu] +  \Gamma_1^{\mu} \cdot\Psi_{\mu} + O(h)  \right )}
\label{eq:effact}
\end{equation}
Computing the corrections involves taking expectation values of the operator $e^{-V/h}$ as well as other operators with respect to the Gaussian functional with covariance $(L^{-1})^{\mu \nu}$, which can be expanded as an infinite series of Gaussian moments.  Fortunately, we can describe the terms in this series graphically using Feynman diagrams.  A result of this analysis is that we can arrange the corrections to the effective action according to the number of loops in the Feynman diagrams, the order in $h$ being given by the number of loops. \cite{zinnjustin}.  $S[a_\mu]$ provides the no loop, or ``tree" level computation.  The lowest order correction that these terms can produce is $O(1)$, which would be an $O(h)$ correction to the effective action (i.e. $1$ loop correction).  This is because the corrections will be given by moments of operators which go as $1/h$ under a Gaussian functional distribution whose covariance goes as $h$.  The terms $\Gamma_1$ and higher produce still higher order corrections.  We discuss in the appendix the connection between the $h$ expansion, which is an artificial parameter, and the effective small parameters in the system, (i.e. the inverse of the distance to criticality and the inverse of the extent of connectivity within the network, as addressed in 
Sec \ref{sec:crit}).

To lowest order we obtain $\Gamma[a_\mu]=S[a_\mu]$ which implies from (\ref{eq:action}) that the equations of motion to lowest order are given by
\begin{eqnarray*}
		\frac{\delta S[a_{\mu}]}{\delta \tilde{a}(x,t)}  &=& \left ( \partial_t + \alpha \right) a(x,t) - f \left ( w * \left [ \tilde{a}(x,t) a(x,t) + a(x,t) \right ] \right)\nonumber \\ 
		&&- \int d^dx'' \, \tilde{a}(x'',t) f^{(1)}\left ( w * \left [ \tilde{a}(x'',t) a(x'',t) + a(x'',t) \right ] \right) w(x''-x) a(x,t) =0\label{eq:classical-phi}\\
		\frac{\delta S[a_{\mu}]}{\delta a(x,t)}  &=& \left ( -\partial_t + \alpha \right) \tilde{a}(x,t) \nonumber\\
		&-& \int d^dx''\,\tilde{a}(x'',t)f^{(1)}\left ( w * \left [ \tilde{a}(x'',t) a(x'',t) + a(x'',t) \right ] \right)w(x'' - x) \left [\tilde{a}(x,t) + 1 \right] =0\nonumber
		\label{eq:classical-phi-tilde}
	\end{eqnarray*}
from which we obtain $\tilde{a}=0$ (because there is no initial condition ``source" term for $\tilde{a}$) and the mean field Wilson-Cowan equation (\ref{eq:wc}).  We can go to higher order by performing a loop expansion on the path integral in (\ref{eq:effact}) and this next correction was computed in \cite{bc}.  

The importance of this approach is that as we consider successive orders in the loop expansion, the effective action closes the system automatically.  If we could calculate $\Gamma[a_\mu]$ for our model of interest, then we would have the exact equation of motion for the true mean of the theory.  In essence, we are trading a closure problem for an approximation problem.  The advantage gained is that we do not have to worry about the contributions of higher moments explicitly and we can consider explicitly the criteria allowing us to truncate the expansion.  If there is an explicit loop expansion parameter, this truncation is straightforward.  If not, as in our case, we must explicitly assess the regimes in which any truncation is valid.  Even in the case where a truncation fails, the loop expansion can provide guidance in terms of identifying classes of diagrams (i.e. terms in the expansion) that are relevant in appropriate limits, which could be summed.

We can now generalize this procedure for equations of motion for an arbitrary number of cumulants by considering a generating functional for an arbitrary number of ``composite operators"~\cite{cornwall}.   
 In the case of the first and second cumulants,  $a(x,t)$ and $C(x,y,t)$, we define  the composite cumulant generating functional
	\begin{eqnarray}
		\exp(W[ J^\mu, K^{\mu\nu} ] )&=&  \int {\cal D}\Phi_{\mu} \exp{\left (-\left [S[\Phi_{\mu}] -  J^\nu(x,t) \cdot\Phi_{\mu}(x,t) \right . \right .} \nonumber \\
		&& \left . \left . - \frac{1}{2}  \Phi_{\mu}(x,t)\cdot  K^{\mu\nu}(x,t;x',t')\cdot \Phi_{\nu}(x',t')  \right]\right ) 
	\end{eqnarray}
We now perform a double Legendre transform to obtain the effective composite action
\begin{eqnarray}
		\Gamma[a_{\mu}, C_{\mu\nu}] &=& -W[J^{\mu}, K^{\mu\nu} ] + J^{\mu}(x,t)\cdot  a_{\mu}(x,t)  \\
		&& + \frac{1}{2}  \left [ a_{\mu}(x,t) a_{\nu}(x',t') + h C_{\mu\nu}(x,t; x',t')\right]\cdot K^{\mu\nu}(x,t;x',t') \nonumber
	\end{eqnarray}
with conditions
\begin{eqnarray}
		\frac{\delta W[J^\mu, K^{\mu\nu}]}{\delta J^\mu(x,t)} &=& a_{\mu}(x,t) \\
		\frac{\delta W[J^{\mu}, K^{\mu\nu}]}{\delta K^{\mu\nu}(x,t;x',t')} &=& \frac{1}{2} \left [ a_{\mu}(x,t) a_{\nu}(x',t') + h C_{\mu\nu}(x,t; x',t')\right ]
	\end{eqnarray}
and
	\begin{eqnarray}
		&&\frac{\delta \Gamma[a_{\mu}, C_{\mu\nu}]}{\delta a_{a}(x,t)} = J^{\mu}(x,t) +\frac{1}{2} a_{\nu}(x',t') \cdot \left [ K^{\mu\nu}(x,t; x',t') + K^{\mu\nu}(x',t'; x,t)\right ] 
		\label{eq:phieom} \\
		&&\frac{\delta \Gamma[a_{\mu}, C_{\mu\nu}]}{\delta C_{\mu\nu}(x,t;x',t')}  = \frac{1}{2}h K^{\mu\nu}(x,t;x',t')
		 \label{eq:geom}
	\end{eqnarray}
	The equations of motion are obtained by setting $J^{\mu}=0$ and $K^{\mu\nu}=0$ in (\ref{eq:phieom}) and (\ref{eq:geom}).  We calculate the equations of motion to lowest order for this system in the appendix.  The results are 
	\begin{eqnarray}
		\left ( \partial_t + \alpha \right) a(x,t) - f \left ( w \star  a(x,t)\right) && \nonumber \\
		- \int d^dx' \, d^dx'' \, f^{(2)}(x,t)w(x-x'')w(x-x')  C_{11}(x',t;x'',t) &=&  0
	\end{eqnarray}
	and for  $C_{\mu\nu}$, we get:
	\begin{eqnarray}
		\left ( \partial_t + \alpha \right) C_{1,-1}(x,t; x_0,t_0) &-& \int d^dx'\, f^{(1)} \left ( w \star  a(x,t)\right)w(x-x') C_{1,-1}(x',t; x_0,t_0) \nonumber \\
		&=& \delta(x-x_0) \delta(t-t_0) \label{eq:flucdis0} \\
		\left (- \partial_t + \alpha \right)C_{-1,1}(x,t;x_0,t_0) &-& \int d^dx'\,f^{(1)} \left ( w \star  a(x',t)\right)w(x'-x) C_{-1,1}(x',t;x_0,t_0)  \nonumber \\
		&=& \delta(x-x_0)\delta(t-t_0) \\
		\left ( \partial_t + \alpha \right) C_{11}(x,t; x_0,t_0) &-& \int d^dx'\, f^{(1)} \left ( w \star  a(x,t)\right)w(x-x') C_{11}(x',t; x_0,t_0)  \nonumber \\
		 - \int d^dx'\,\left [ f^{(1)} \left ( w \star  a(x,t)\right) w(x-x') a(x',t') \right . &+& \left . f^{(1)} \left ( w \star  a(x',t)\right) w(x'-x) a(x,t) \right ] C_{-1,1}(x',t;x_0,t_0) \nonumber \\
		 &=&0 \label{eq:flucdis} 
	\end{eqnarray}
	together with the conditions
	\begin{eqnarray}
		C_{11}( x,t; x',t') &=& C_{11}( x',t'; x,t) \\
		C_{-1,1}( x,t; x',t') &=& C_{1,-1}( x',t'; x,t)
	\end{eqnarray}
	The 2-point correlation designated as $C_{\mu\nu}(x,t;x',t')$ generalizes the cumulant $C(x,y,t)$ in (\ref{eq:2ndnoc}) to include  both the unequal time 2-point correlation ($C_{11} $) and the linear response ($C_{1,-1}$).  The equation for $C(x,y,t)$ is  obtained by taking the equation for $\lim_{t\rightarrow t_0} C_{11}(x,t; x_0, t_0) + C_{11}(x_0,t_0; x,t)$ which results in (\ref{eq:genConWcC}).
	Note that we have also produced an equation for the Green's function of the theory $C_{1,-1}$ as well as its time reversed counterpart.   Time reversal involves swapping the fields $\varphi$ and $\tilde{\varphi}$.  In the time reversed theory, $\tilde{\varphi}$ plays the role of activity.  Time reversal does not give an equivalent theory since our Markov process is not time reversal invariant in general.
In field theory language, this system of equations is known as the 2PI equations and we adopt this moniker for convenience\footnote{``2PI" stands for 2 Particle Irreducible.  The effective action $\Gamma[a_\mu]$ is the generating functional of 1PI graphs, which means that the graphs which determine $\Gamma[a_\mu]$ cannot be disconnected by cutting any single line of the graph.  Similarly, $\Gamma[a_\mu, C_{\mu \nu}]$ is 2PI is the sense that graphs contributing to it cannot be disconnected by cutting two lines.}.  We can then analogously derive nPI equations for any number of normal ordered cumulants.

With the moment hierarchy approach, in order to produce better approximations we are required to truncate further in the hierarchy.  This can quickly produce unwieldy equations.  The loop expansion provides an alternative in that corrections to the generalized equations can be produced with a diagrammatic expansion, namely the one which calculates $\Gamma_2[a_{\mu}, C_{\mu\nu}]$, from which the corrections to the equations can be calculated.



\section{All-to-All Networks, Finite Size Effects, and Simulations}
\label{sec:alltoall}

We consider the example of an all-to-all system, wherein each neuron connects to the entire network. Mean field theory should work well in this case because the number of post-synaptic neurons reduces the coupling of the fluctuations.  In this case, the fluctuations reasonably reduce to corrections due to the finite size of the network, as we would expect.  We take the weight function to be a constant, normalized by the number of neurons in the system $w_{ij} = w_0/N$
for some $w_0$.  The generalized Wilson-Cowan equations become
\begin{eqnarray}
	\partial_t a_i(t) &=& - \alpha a_i(t) + f \left ( w_0 \frac{1}{N} \sum_j a_j(t) + I_i(t) \right) \nonumber \\
	&&+ \frac{1}{2} f''\left ( w_0 \frac{1}{N} \sum_j a_j(t) + I_i(t) \right ) w_0^2 \frac{1}{N^2}\sum_{jk}  C_{jk}(t)
\end{eqnarray}

\begin{eqnarray}
	\frac{d}{dt}C_{ij}(t) &=& -2\alpha C_{ij}(t) + f' \left ( s_i \right ) w_0 \frac{1}{N} \sum_{k} C_{kj}(t) + f'\left ( s_j \right ) w_0 \frac{1}{N}\sum_{k}  C_{ik}(t) \nonumber \\
	&&+ \frac{1}{N} \left ( f' \left ( s_i \right ) w_0 a_j(t) + f' \left ( s_j \right ) w_0 a_i(t)  \right )
\end{eqnarray}
We can simplify this further by taking the initial conditions $a_i(0) = a_0$ and $C_{ij}(0) = 0$  and assuming homogeneous external input $I$.  This corresponds to initial conditions in the network determined by a Poisson distribution with mean $a_0$. $C_{ij}(0)=-a_i(0)\delta_{ij}$ would indicate starting with precisely $a_i(0)$ active neurons at $i$ at time $t=0$. Then symmetry reduces the equations to
\begin{eqnarray}
	\partial_t a(t) &=& - \alpha a(t) + f \left ( w_0 a(t) + I(t) \right) \nonumber \\
	&&+ \frac{1}{2} f''\left ( w_0 a(t) + I(t) \right ) w_0^2  C(t)
	\label{eq:aAA}
\end{eqnarray}

\begin{eqnarray}
	\frac{d}{dt}C(t) &=& -2\alpha C(t) + 2 f' \left ( w_0 a(t) + I(t) \right ) w_0   C(t)  \nonumber \\
	&&+ \frac{1}{N}2 f' \left ( w_0 a(t) + I(t) \right ) w_0 a(t) 
	\label{eq:CAA}
\end{eqnarray}
where we have defined 
\begin{eqnarray}
	a(t) &=& \frac{1}{N} \sum_i a_i(t) \\
	C(t) &=& \frac{1}{N^2} \sum_{ij} C_{ij}(t)
\end{eqnarray}
Note that as $N \rightarrow \infty$ the source term for $C(t)$ vanishes, which implies that $C(t)$ decouples from the equation for $a(t)$, which then reduces to the standard Wilson-Cowan equation.  The matrix $\Gamma_{ij}$ in this case is the function
\begin{equation}
	\Gamma[a^0] =  \alpha - f' \left ( w_0 a^0 + I(t) \right) w_0
\end{equation}
The steady state value of $C(t)$ is given by
\begin{equation}
	C(t) =  \frac{f'(s_0)w_0a_0}{N\Gamma[a_0]}
\end{equation}
The relative size of the fluctuations in steady state is determined by the product $N\Gamma[a_0]$.  Large networks or networks distant from a bifurcation experience reduced correlations.

\begin{figure}
\begin{center}
\scalebox{0.25}{\includegraphics{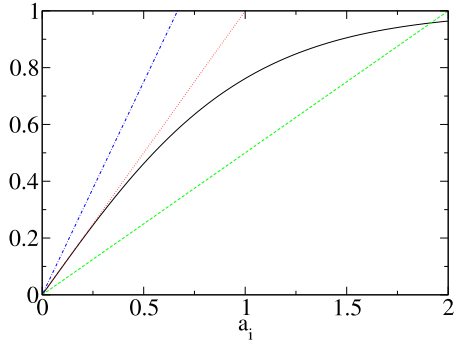}}
\end{center}
\caption{Graphical depiction of solutions to equation~(\ref{eq:meanFieldBif}) for various values of $\alpha$.  }
\label{fig:meanFieldBif}
\end{figure}

\begin{figure}

 \begin{minipage}{8.0cm} \centering{\scalebox{0.25}{\includegraphics{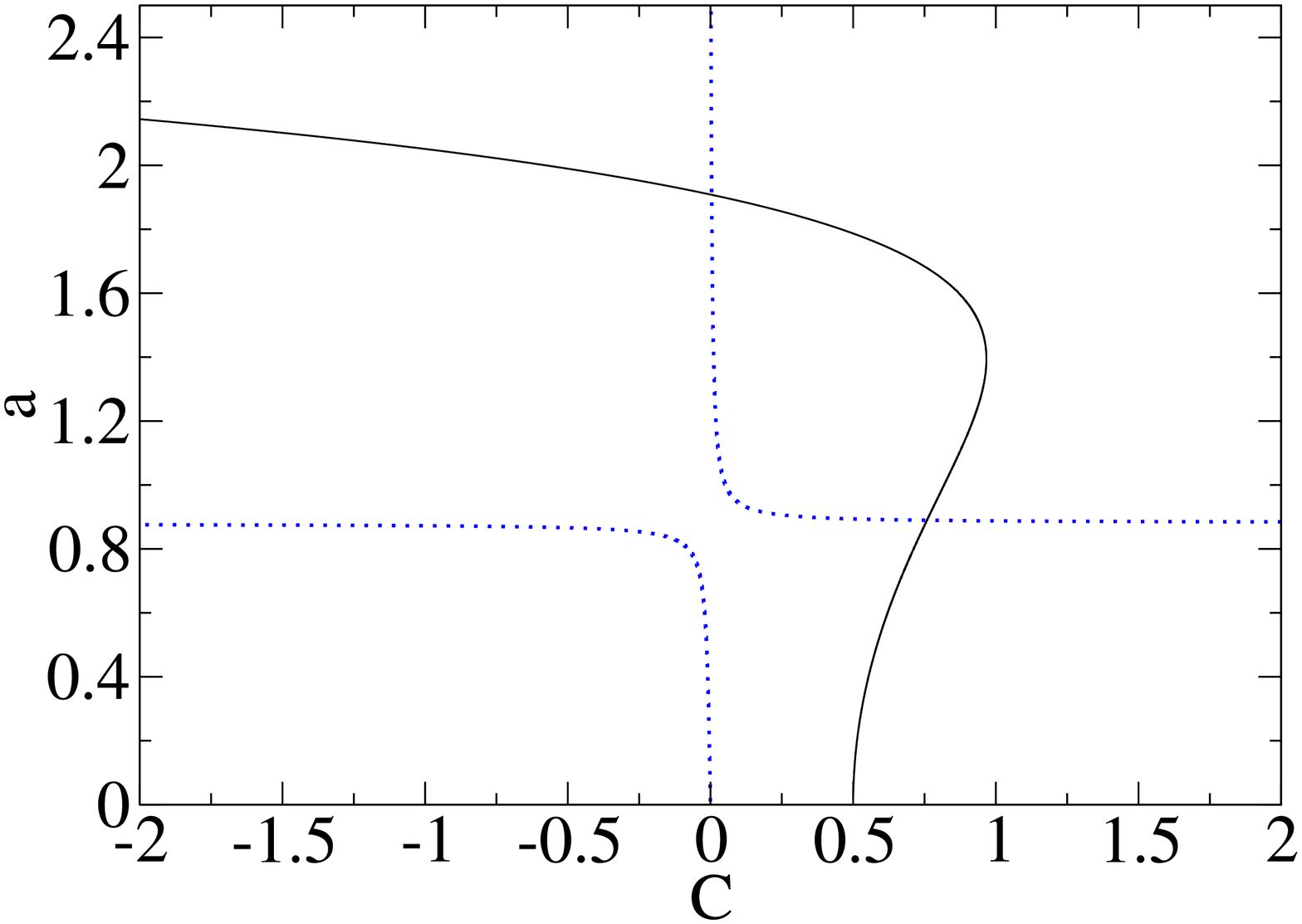}} (a)}
 \end{minipage}
 \vspace{1.0cm}
 \begin{minipage}{8.0cm} \centering{\scalebox{0.25}{\includegraphics{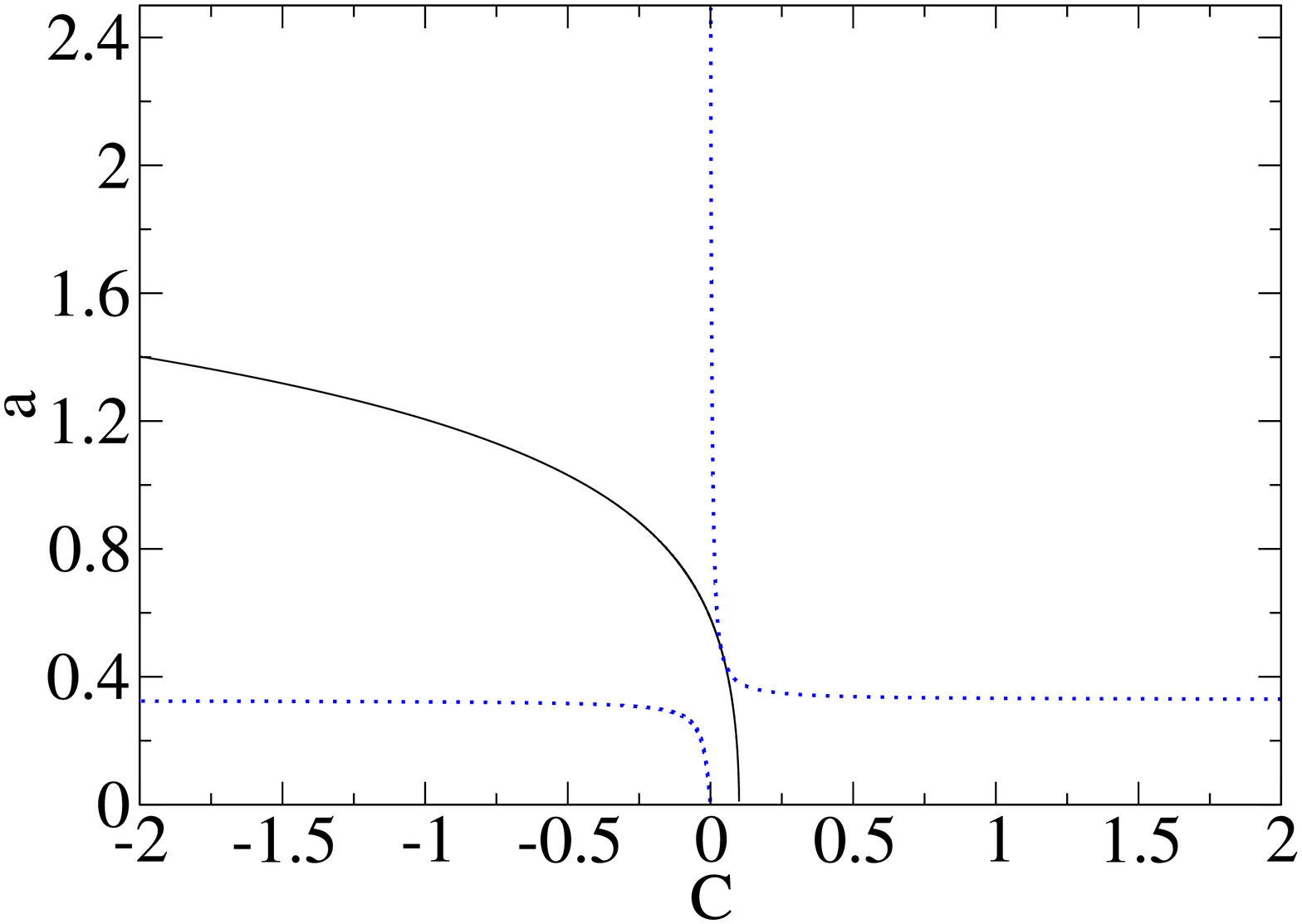}} (b)}
 \end{minipage}
 \begin{minipage}{8.0cm}
\centering{ \scalebox{0.25}{\includegraphics{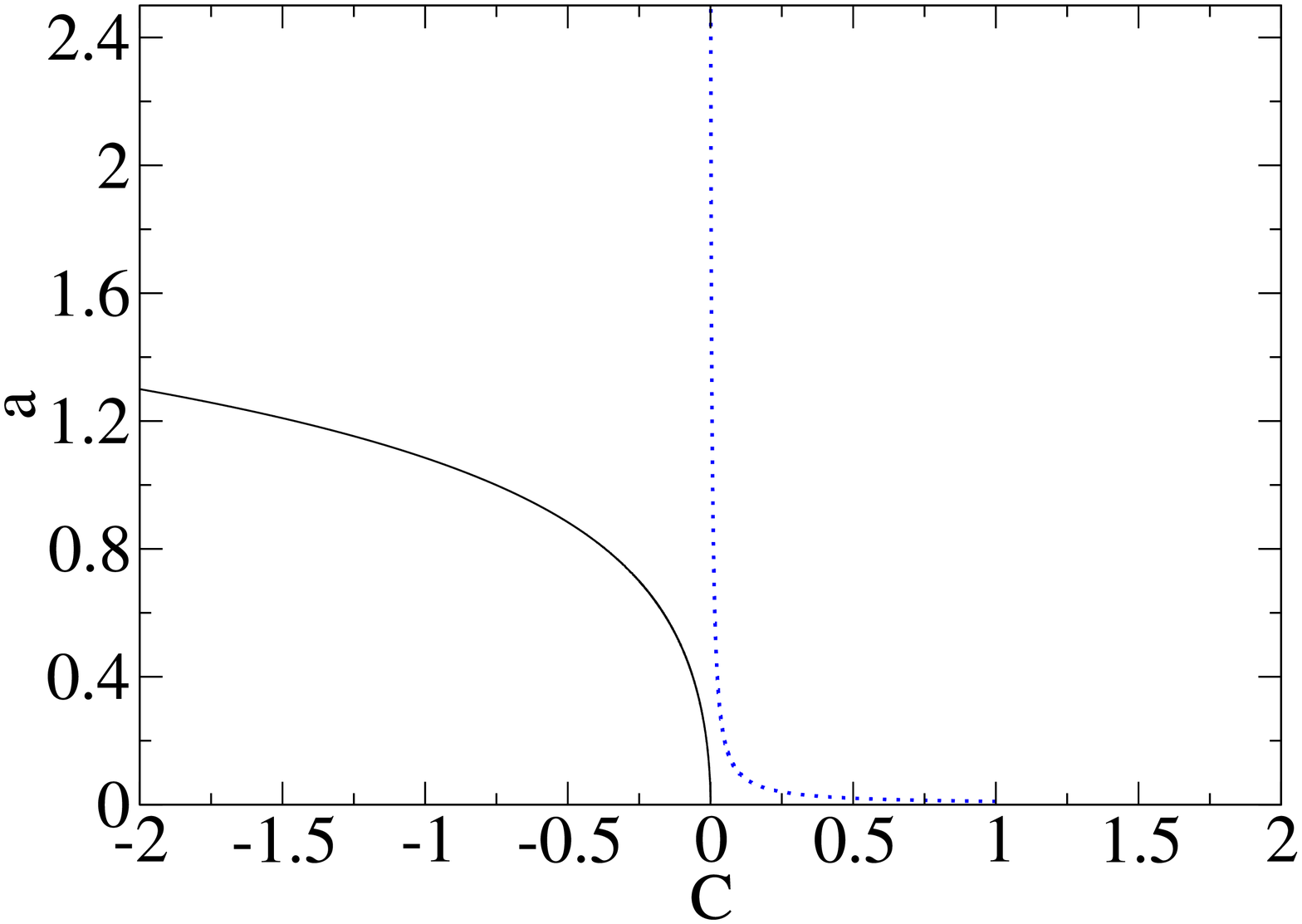}} (c) }
\end{minipage}
 \caption{Phase planes for the all-to-all generalized equations with a) $\alpha= 0.5$, b) $\alpha = 0.9$, and c) $\alpha=1.0$.  $N=100$.  Solid (black) lines are $a$ nullclines; dotted (blue) lines are $C$ nullclines.}
 \label{fig:phasePlane1}
 \end{figure}

\begin{figure}
 \begin{minipage}{8.0cm} \scalebox{0.25}{\includegraphics{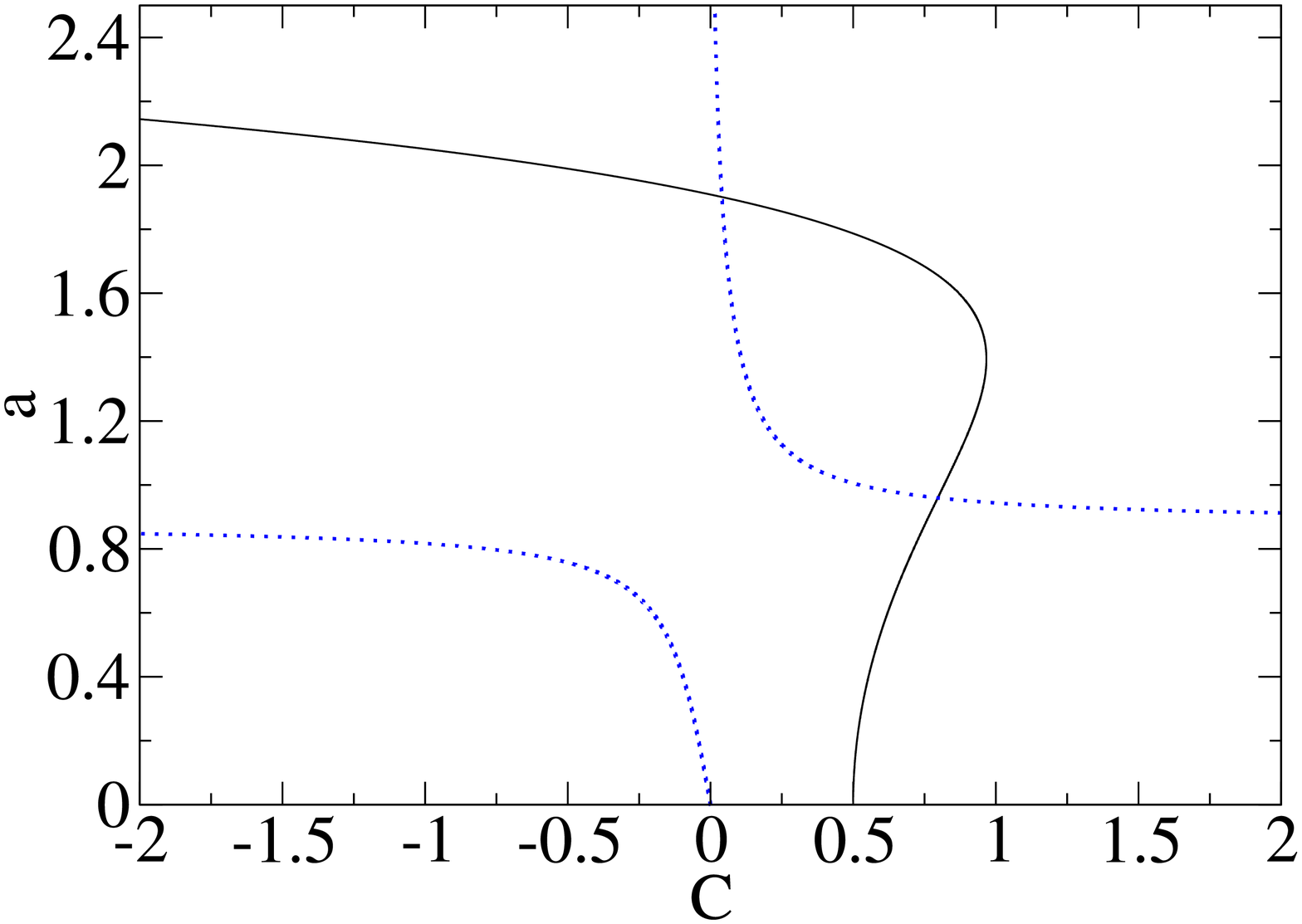}} (a) \end{minipage}
 \vspace{1.0cm}
 \begin{minipage}{8.0cm} \scalebox{0.25}{\includegraphics{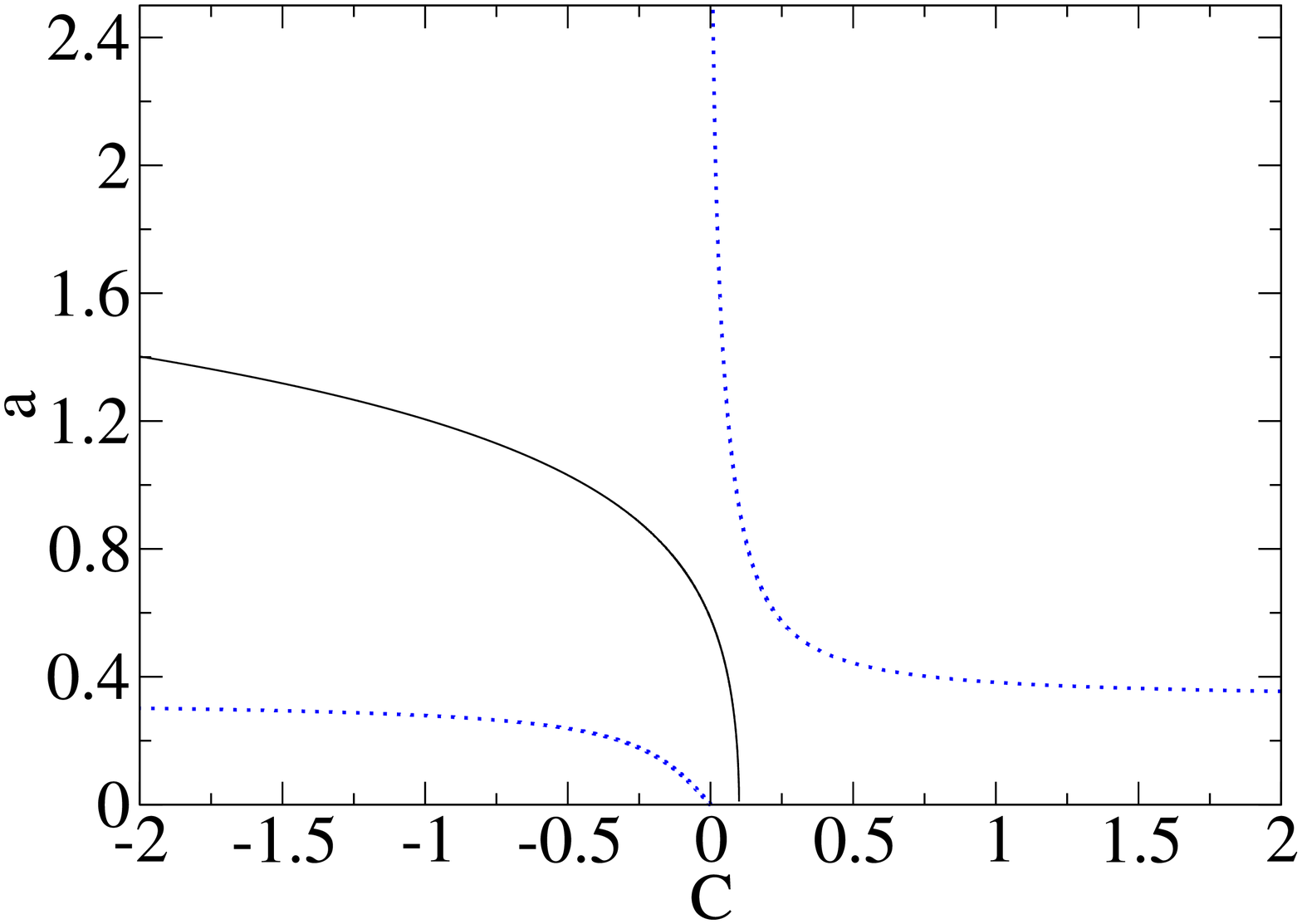}} (b) \end{minipage}
 \begin{minipage}{8.0cm} \scalebox{0.25}{\includegraphics{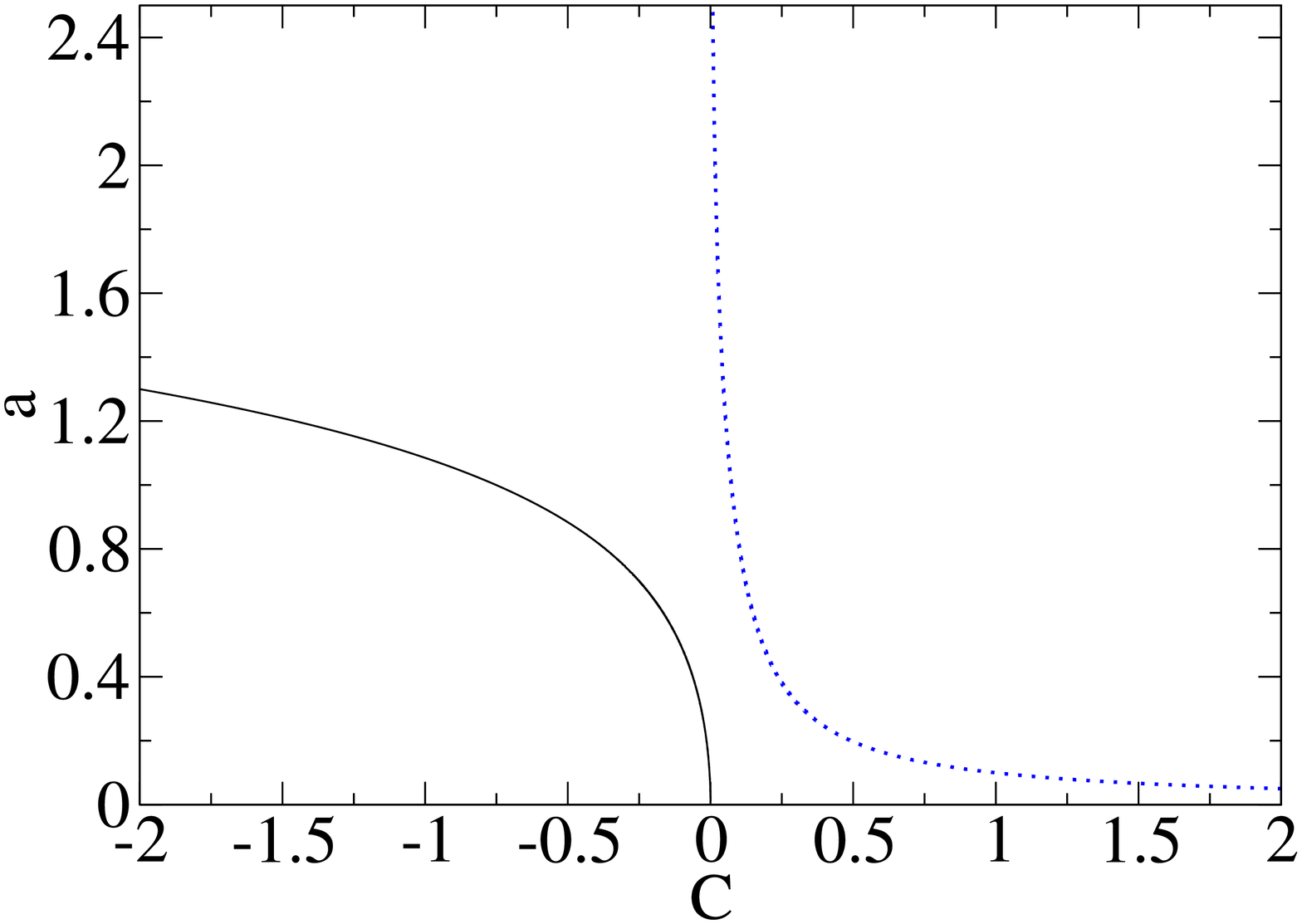}} (c) \end{minipage}
 \caption{Phase planes for the all-to-all generalized equations with $\alpha= 0.5$ on the left, $\alpha = 0.9$ in the center, and $\alpha=1.0$ on the right.  $N=10$.  Solid (black) lines are $a$ nullclines; dotted (blue) lines are $C$ nullclines.}
 \label{fig:phasePlane2}
 \end{figure}

We now examine the phase plane of this simplified system.  For concreteness, consider the firing rate function $f(s)$ to be
\begin{equation}
	f(s) = \tanh(s)\Theta(s)
	\label{eq:sigfunc}
\end{equation}
where $\Theta$ is the Heaviside step function.  At mean field level (i.e. consider $C(t)$ to be zero) equilibria are determined by solutions of the equation
\begin{equation}
	\alpha a_0 = f(a_0)
	\label{eq:meanFieldBif}
\end{equation}
Figure~\ref{fig:meanFieldBif} graphically displays the solutions of this equation.
From the figure, we see that  the equation exhibits a  bifurcation as the value of $\alpha$ (the slope of the straight line in the figure) is decreased.  The critical value for this bifurcation is $\alpha = 1.0$.  We will refer to the non zero stable equilibrium as the ``activated" state.  The generalized activity equations (\ref{eq:aAA}) and (\ref{eq:CAA}) also exhibit a bifurcation.  The phase plane is shown in Figure~\ref{fig:phasePlane1} for $\alpha = 0.5, 0.9$ and $1.0$ with $N=100$ and in Figure~\ref{fig:phasePlane2} for $N=10$ with the same values of $\alpha$.  As expected, the steady state value of $C$ is larger for $N=10$.  Note that the nullclines for $C(t)$ display divergences associated with $\Gamma$ approaching zero.  In addition note that the location of the bifurcation is different.  For $N=100$ the bifurcation happens near $\alpha = 0.9$ and with $N=10$ the bifurcation happens for  $0.9 < \alpha < 1.0$.  Because the generalized equations are a coupled system, it is possible that more interesting bifurcation structure may be manifested.
In addition to the fixed points which exist in mean field theory there is a new fixed point.  Whereas the bifurcation at mean field level is a pitchfork bifurcation, that of the generalized equations is a saddle node bifurcation, with an unstable fixed point emerging as the $a$ and $C$ nullclines cross.  There is always a fixed point at $a = 0$ and $C=0$ because it is an absorbing state.

Importantly, we see that we can alter the bifurcation structure by adding a forcing or source term to the correlation function $C(t)$ equation, linearly shifting the $C$ nullcline.  This removes the stable fixed point for high $a$ (the one associated with the activated state in mean field theory).  Because of this, we see that we can disrupt the activated state by stimulating the system with an input such that the correlation is sourced more strongly than the mean.   We can use this to ``turn off" the activated state by synchronizing the network.  These correlations drive the system to the absorbing state of the full model. To reverse this deactivation, we simply drive the system with Poisson noise (i.e. force the equation for $a(t)$ but not $C(t)$).  Compare this to the effect demonstrated in \cite{laingchow} in which synchronized activity associated with fast synapses led to the destabilization of activity which the Wilson Cowan equation predicts to be stable.  For a saturating firing rate function (more generally a function such that $f''(s) < 0$ in the appropriate region) increased correlations inhibit the mean activity $a_i(t)$.

\begin{figure}
\begin{minipage}{8.0cm} \scalebox{0.25}{\includegraphics{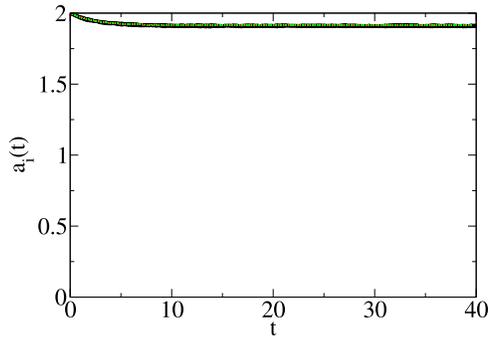}} (a) \end{minipage}
\vspace{2.0cm}
\begin{minipage}{8.0cm} \scalebox{0.25}{\includegraphics{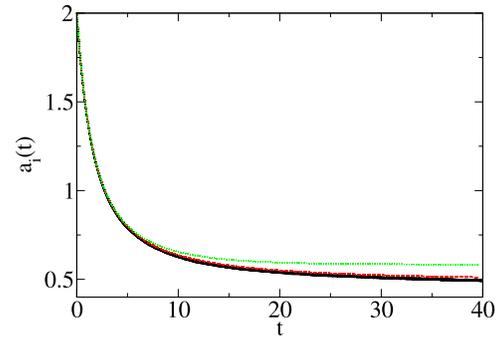}} (b) \end{minipage}
\begin{minipage}{8.0cm} \scalebox{0.25}{\includegraphics{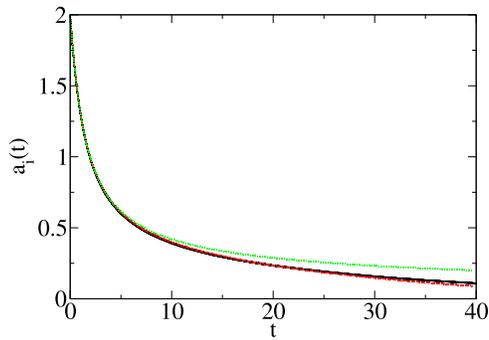}} (c) \end{minipage}
\caption{$a(t)$ vs. $t$ for a)  $\alpha = 0.5$, b) $\alpha = 0.9$, and c) $\alpha = 1.0$.  $N=100$.  Dotted (green) lines are solutions of mean field theory.  Dashed (red) lines are solutions of the generalized equations.  Solid (black) lines are expectations values of data from simulations of the Markov process.}
\label{fig:sim100a}
\end{figure}

\begin{figure}
\begin{minipage}{8.0cm} \scalebox{0.25}{\includegraphics{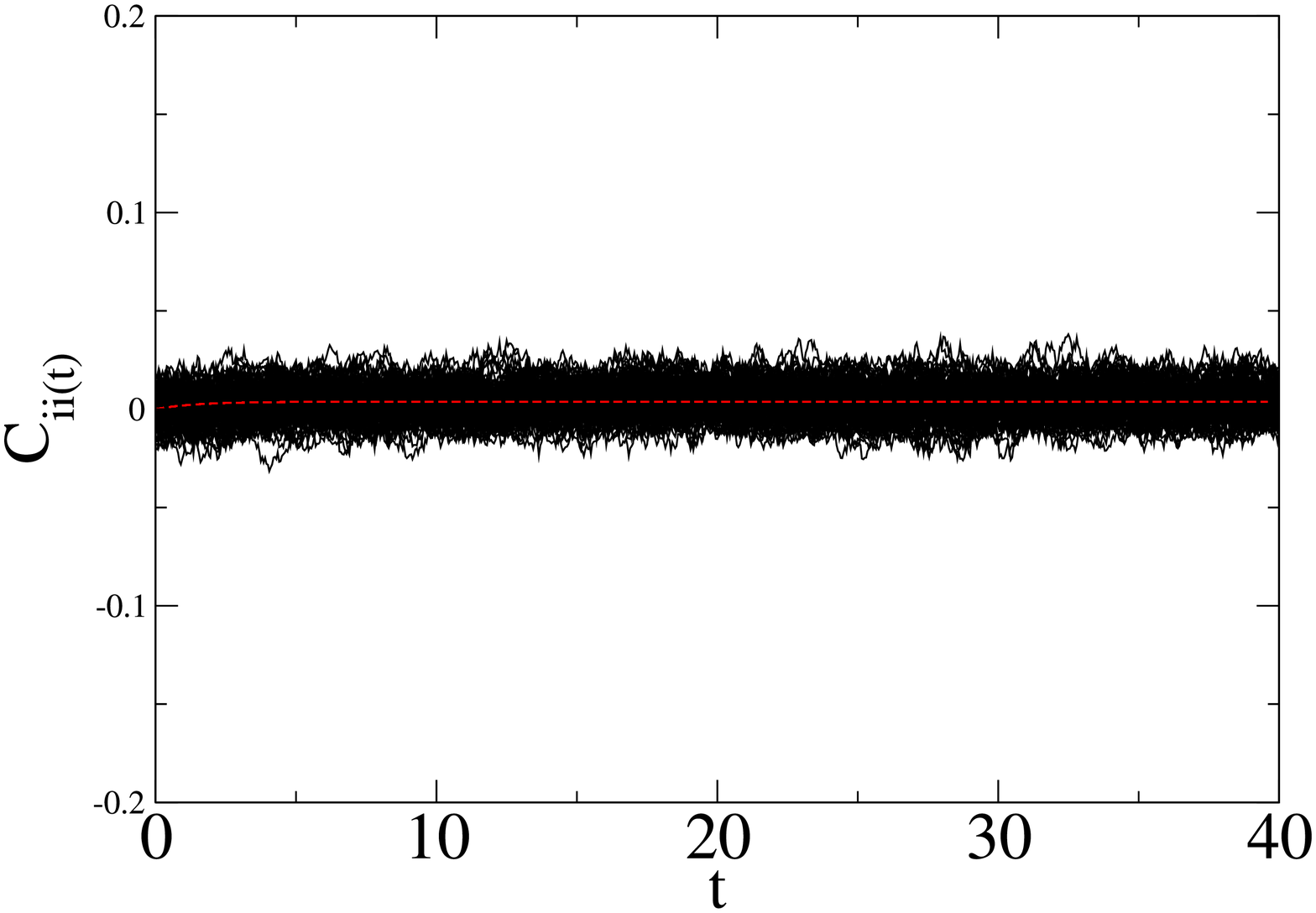}} (a) \end{minipage}
\vspace{1.0cm}
\begin{minipage}{8.0cm} \scalebox{0.25}{\includegraphics{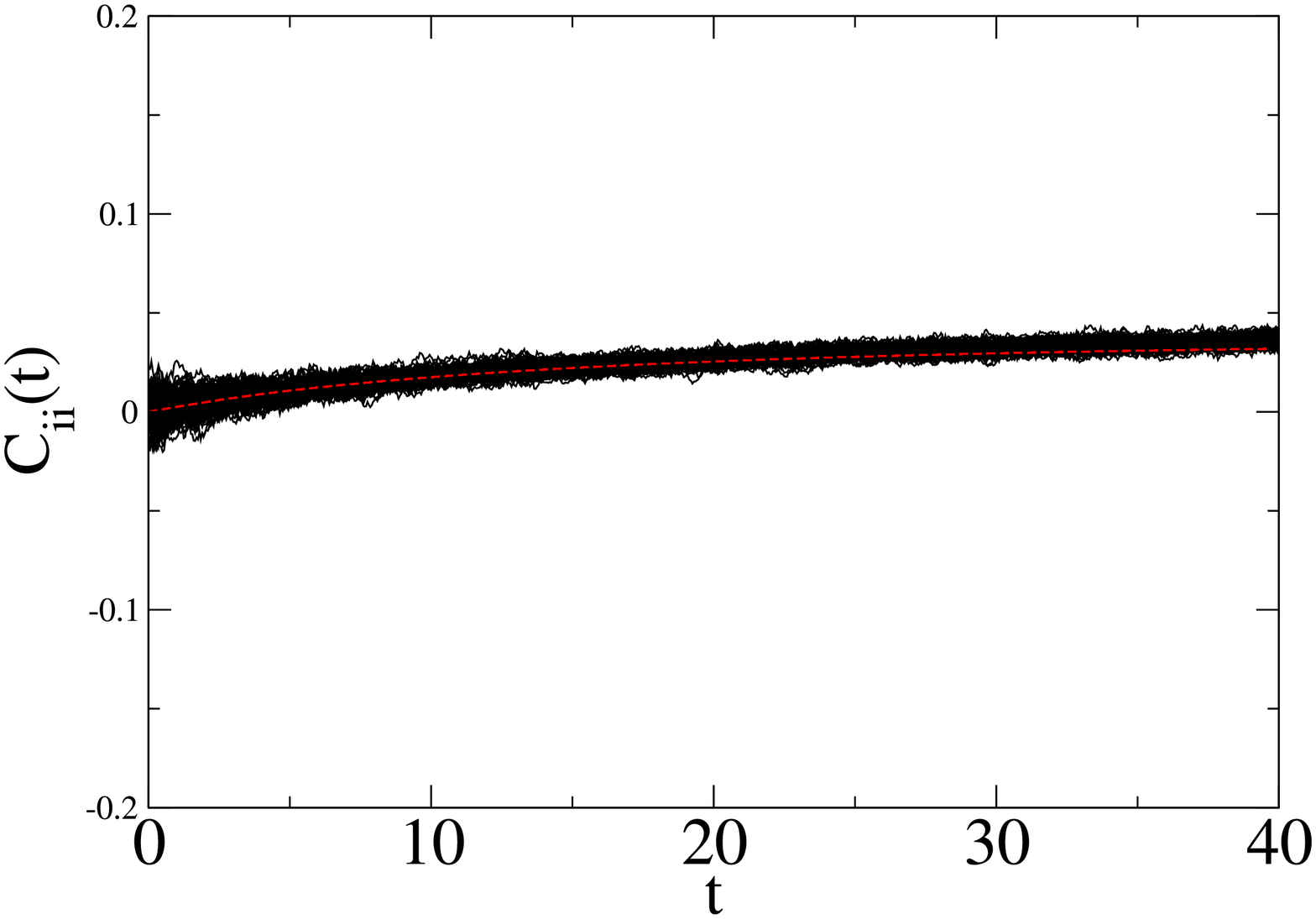}} (b) \end{minipage}
\begin{minipage}{8.0cm} \scalebox{0.25}{\includegraphics{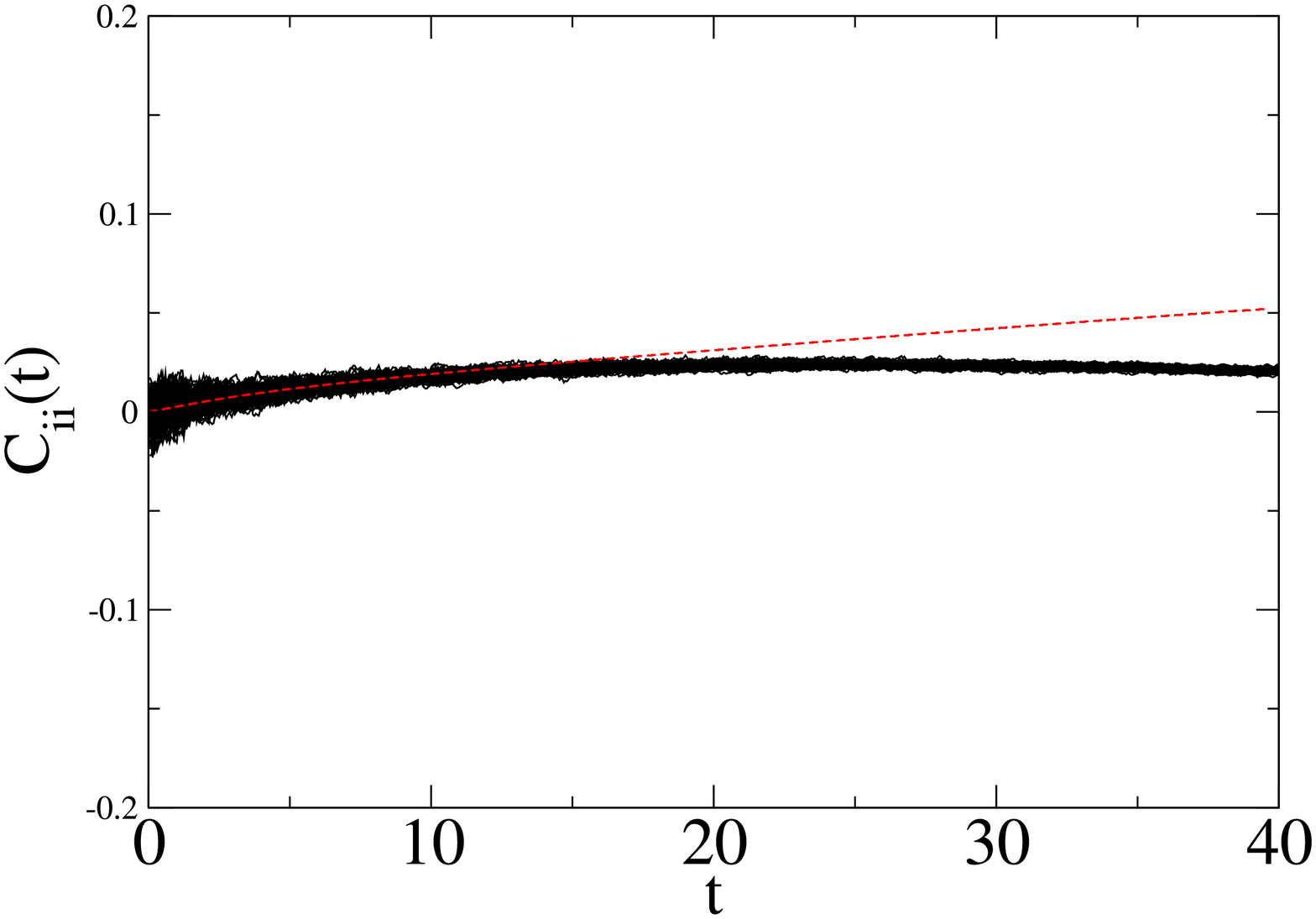}} (c) \end{minipage}
\caption{$C(t)$ vs. $t$ for a)  $\alpha = 0.5$, b) $\alpha = 0.9$, and c) $\alpha = 1.0$. $N=100$. Dashed (red) lines are solutions of the generalized equations.  Solid (black) lines are expectations values of data from simulations of the Markov process.}
\label{fig:sim100C}
\end{figure}

\begin{figure}
\begin{minipage}{8.0cm} \scalebox{0.25}{\includegraphics{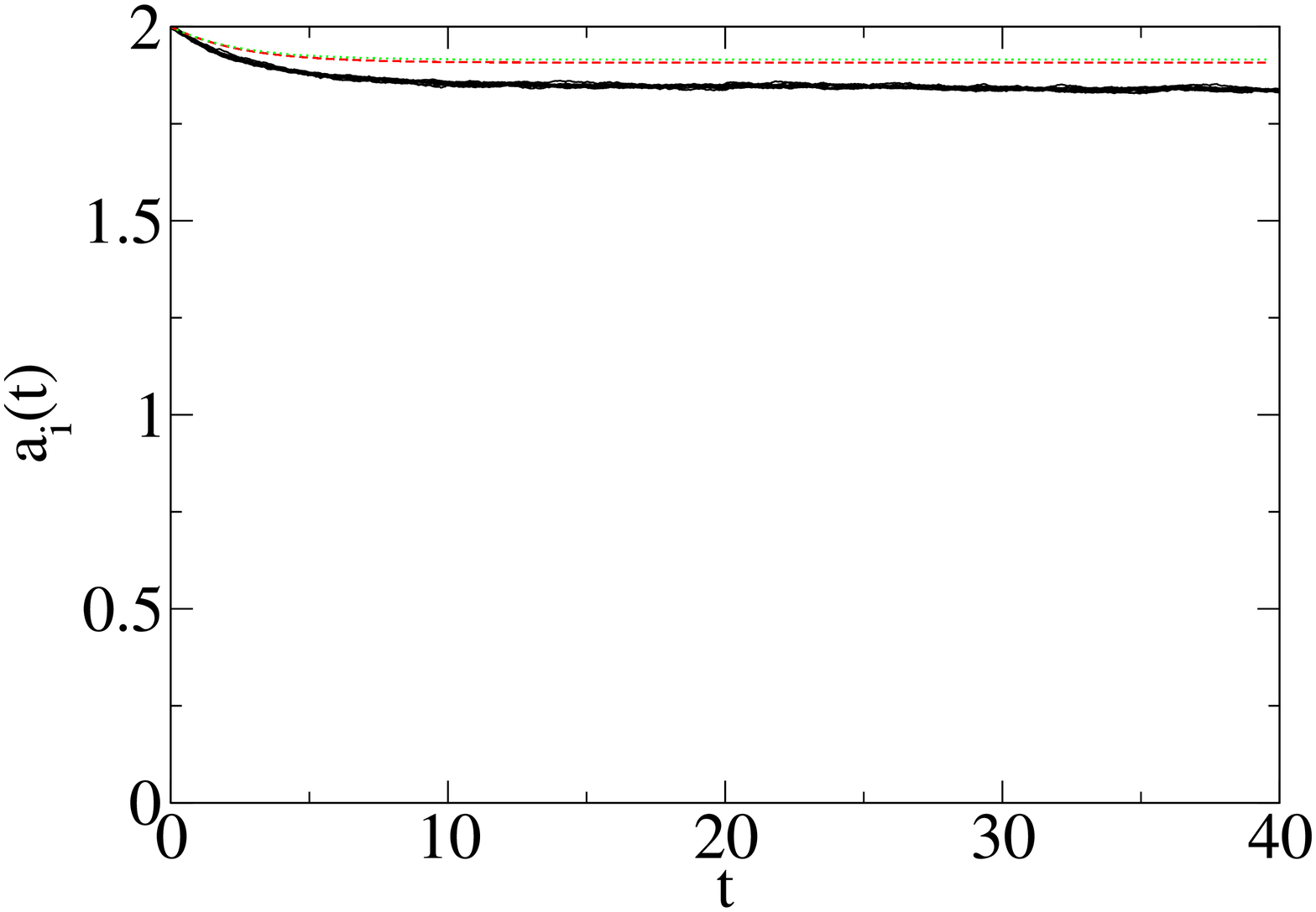}} (a) \end{minipage}
\vspace{1.0cm}
\begin{minipage}{8.0cm} \scalebox{0.25}{\includegraphics{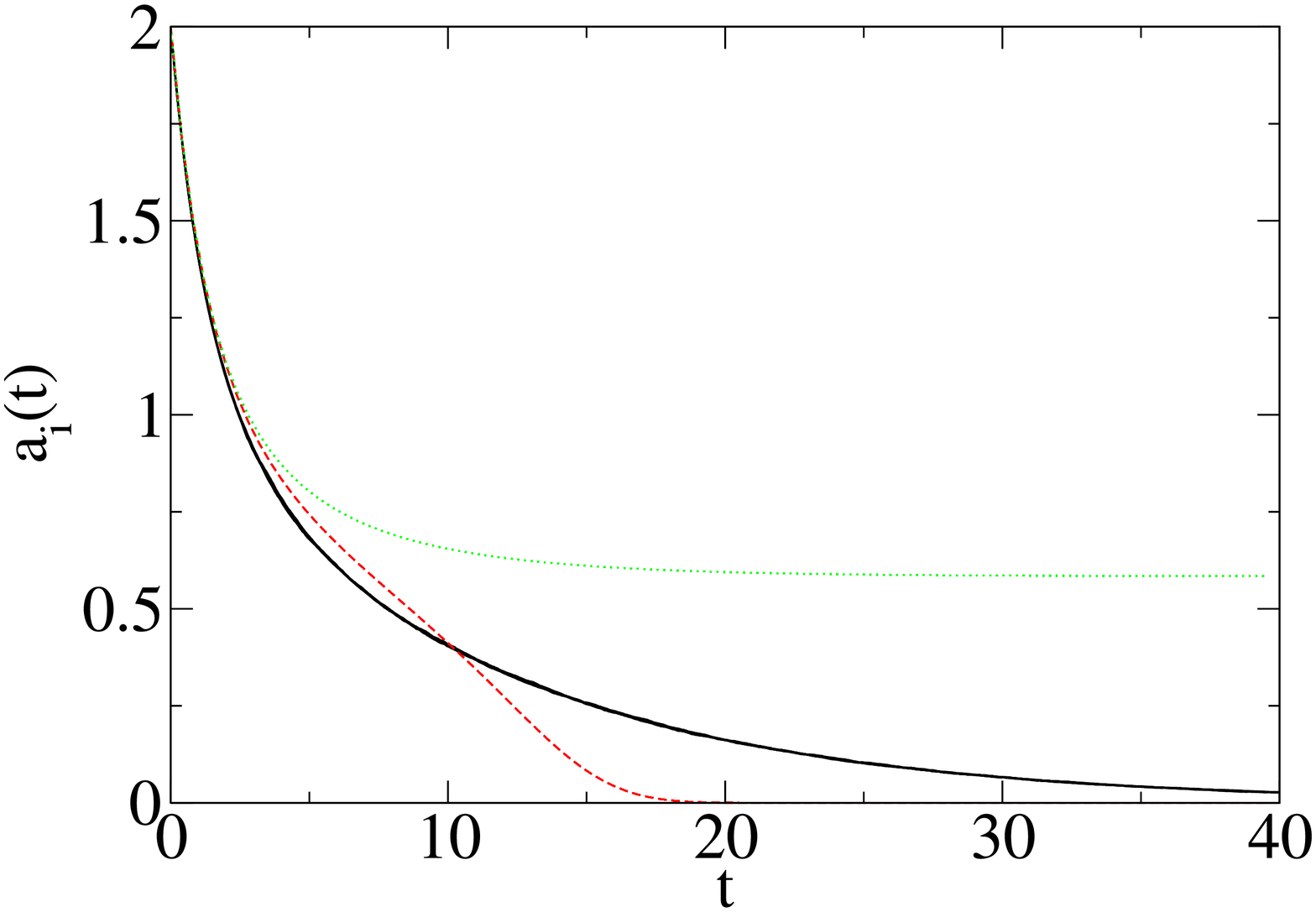}} (b) \end{minipage}
\begin{minipage}{8.0cm} \scalebox{0.25}{\includegraphics{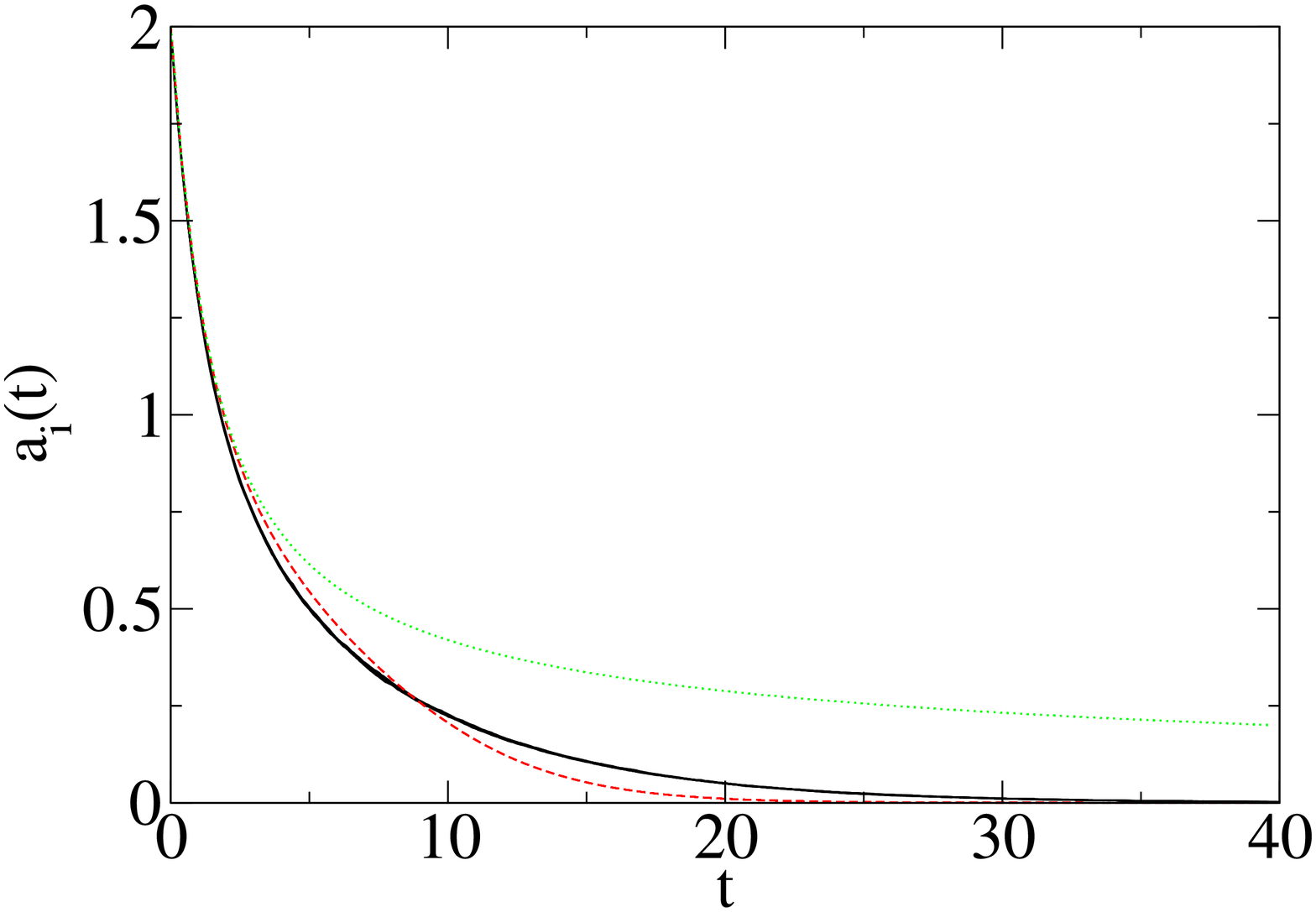}} (c) \end{minipage}
\caption{$a(t)$ vs. $t$ for a)  $\alpha = 0.5$, b) $\alpha = 0.9$, and c) $\alpha = 1.0$. $N=10$. Dotted (green) lines are solutions of mean field theory.  Dashed (red) lines are solutions of the generalized equations.  Solid (black) lines are expectations values of data from simulations of the Markov process.}
\label{fig:sim10a}
\end{figure}

\begin{figure}
\begin{minipage}{8.0cm} \scalebox{0.25}{\includegraphics{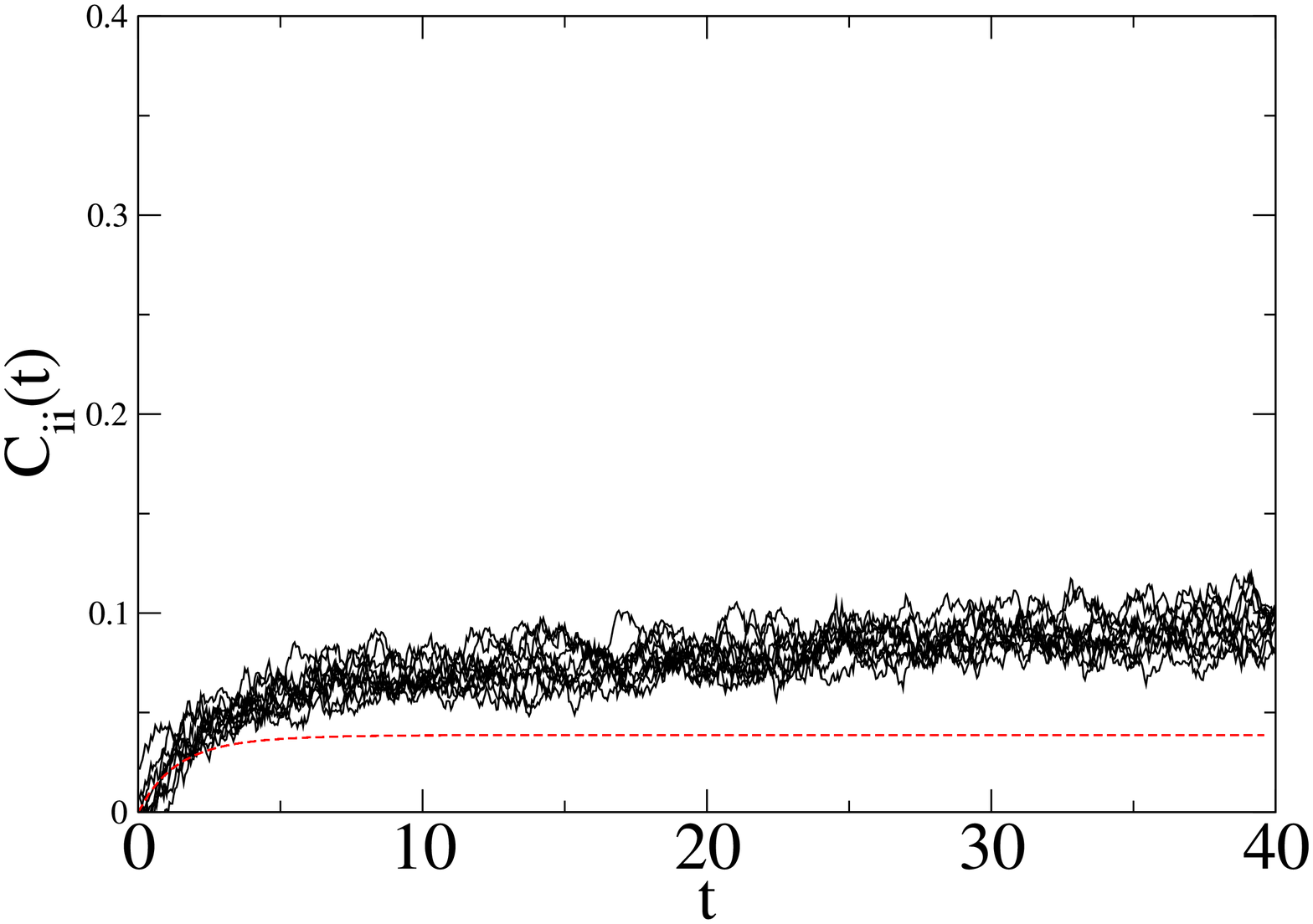}} (a) \end{minipage}
\vspace{1.0cm}
\begin{minipage}{8.0cm} \scalebox{0.25}{\includegraphics{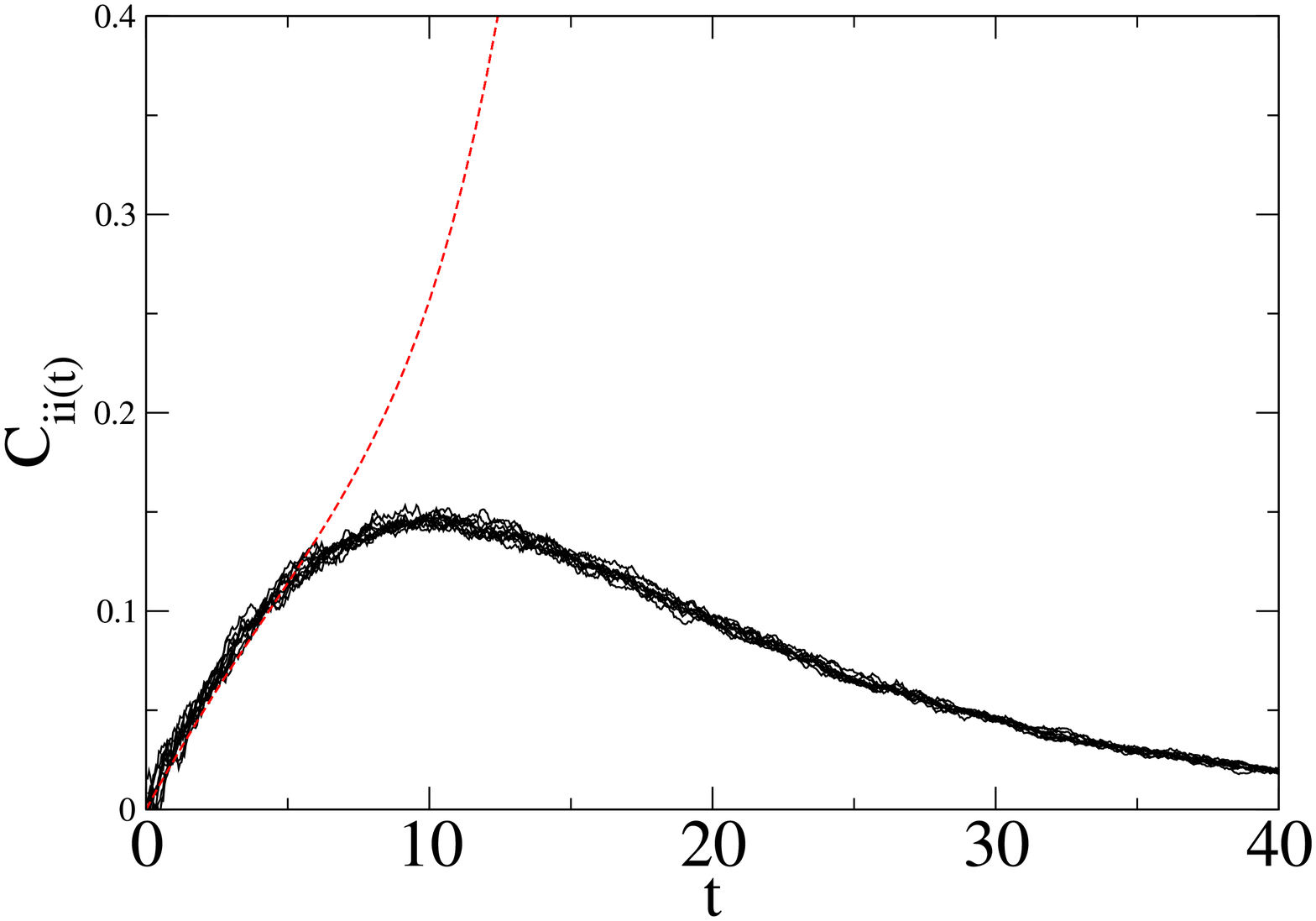}} (b) \end{minipage}
\begin{minipage}{8.0cm} \scalebox{0.25}{\includegraphics{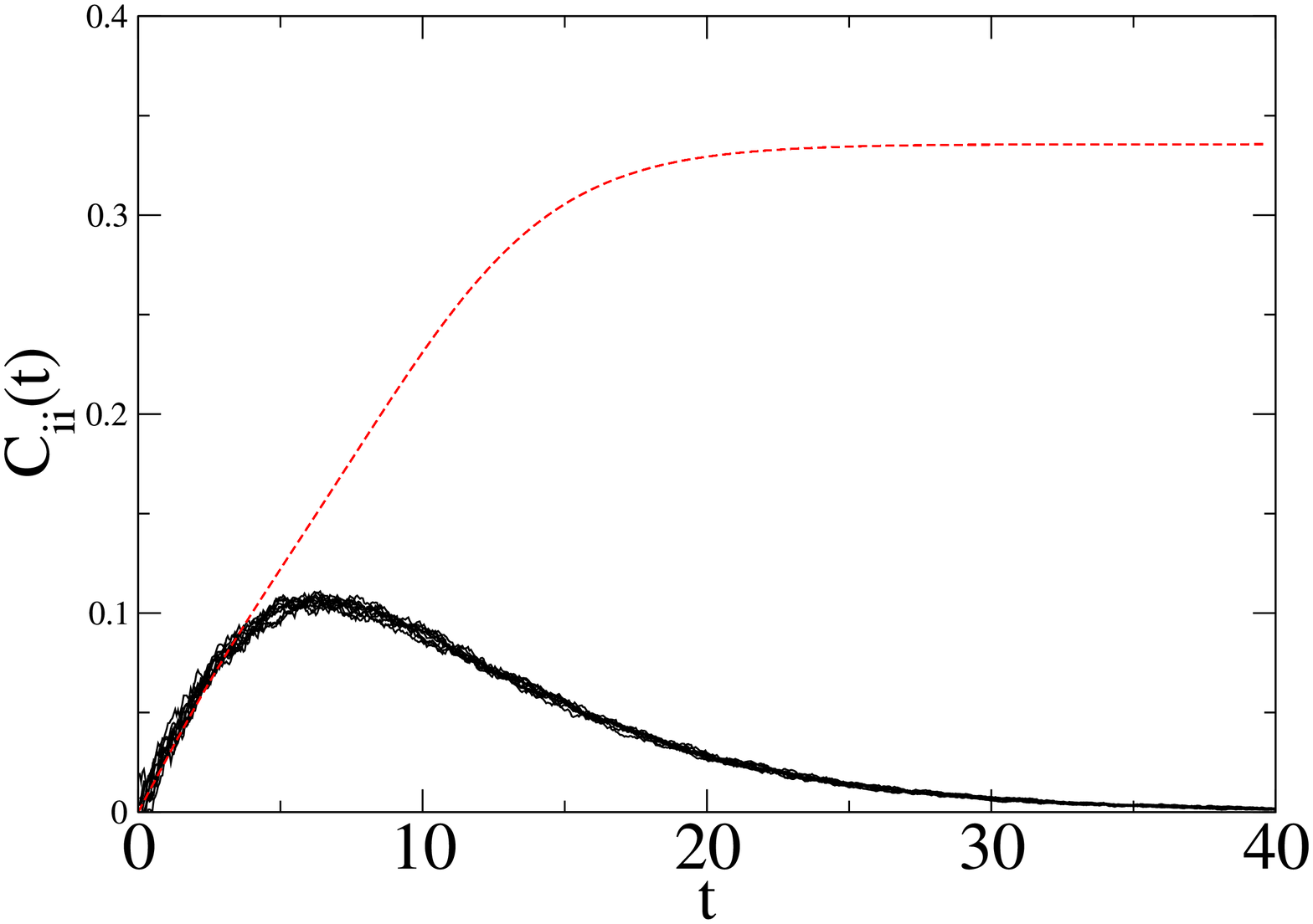}} (c) \end{minipage}
\caption{$C(t)$ vs. $t$ for a)  $\alpha = 0.5$, b) $\alpha = 0.9$, and c) $\alpha = 1.0$. $N=10$.  Dashed (red) lines are solutions of the generalized equations.  Solid (black) lines are expectations values of data from simulations of the Markov process.}
\label{fig:sim10C}
\end{figure}

We now demonstrate the utility of the generalized activity equations (\ref{eq:genDisWc}) and (\ref{eq:genWcC})  for describing the full Markov system (\ref{eq:master}) away from a bifurcation point.  
In order to simulate the Markov model we use a form of the Gillespie algorithm and take expectation values over many time evolutions of the system. We use the firing rate function (\ref{eq:sigfunc}).  It is important to remember when comparing results with simulations that we use the $F(\vec{n})$ whose normal ordered form is $f(s)$.  In this all-to-all case, we need only consider the correction  arising from the quadratic portion of the firing rate function since the corrections will go as powers of the weight function, which in this all-to-all case means they carry factors of $1/N$.  In particular, to lowest order we have
\begin{equation}
F(s) = f(s)
	-f''(s)\frac{1}{N}w_0^2 a + \cdots
\end{equation}
We plot $a_i(t)$ and $C_{ii}(t)$ versus $t$ for various values of $\alpha$ and $N$ in Figures~\ref{fig:sim100a} through \ref{fig:sim10C}.  (Note that we are numerically evaluating the generalized equations, not the simplified equations in (\ref{eq:aAA}) and (\ref{eq:CAA}), and comparing them with Monte Carlo simulations of the Markov system; the plots overlay data for each of the $N$ neurons.) We initialize the network with Poisson initial conditions: $a_i(0) = 2$ and $C_{ij}(0) = 0$.  The simulations of the Markov process are averaged over $10^5$ instances.  One can see that the equations match the simulations quite well away from the critical point.  As one approaches the bifurcation, however, the simulations begin to deviate.  At $\alpha = 0.5$, mean field and the generalized equations each match the simulated Markov process.  As one approaches the mean field bifurcation point of $\alpha = 0.9$, the mean field equations no longer match well, but the generalized equations account for the deviation.  From Figure~\ref{fig:sim100C}, one can see that as we approach $\alpha = 1.0$, the estimate of the correlation from the generalized equations becomes poorer.

The plots for $N=10$ in Figures~\ref{fig:sim10a} and \ref{fig:sim10C} demonstrate the breakdown of the generalized equations.  There is already a significant deviation of both mean field and the generalized equations at $\alpha = 0.5$.  Naturally, the discrepancy is accounted for by the poor estimation of the correlation at this level.  As we near $\alpha = 1.0$, the estimate of the correlations begins to grow, whereas the simulated correlation is dropping to zero.

\begin{figure}
\begin{center}
\scalebox{0.25}{\includegraphics{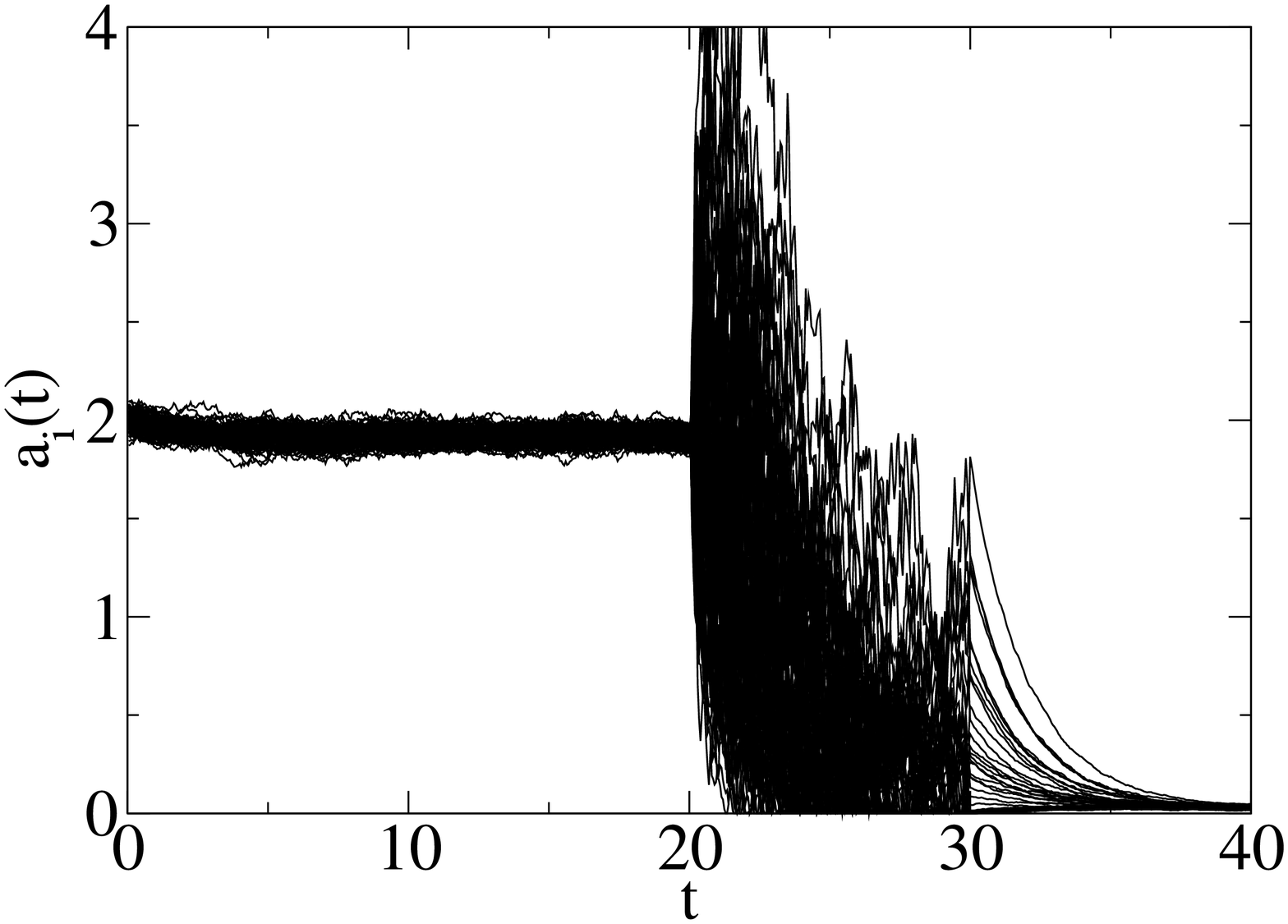}} (a) 
\scalebox{0.25}{\includegraphics{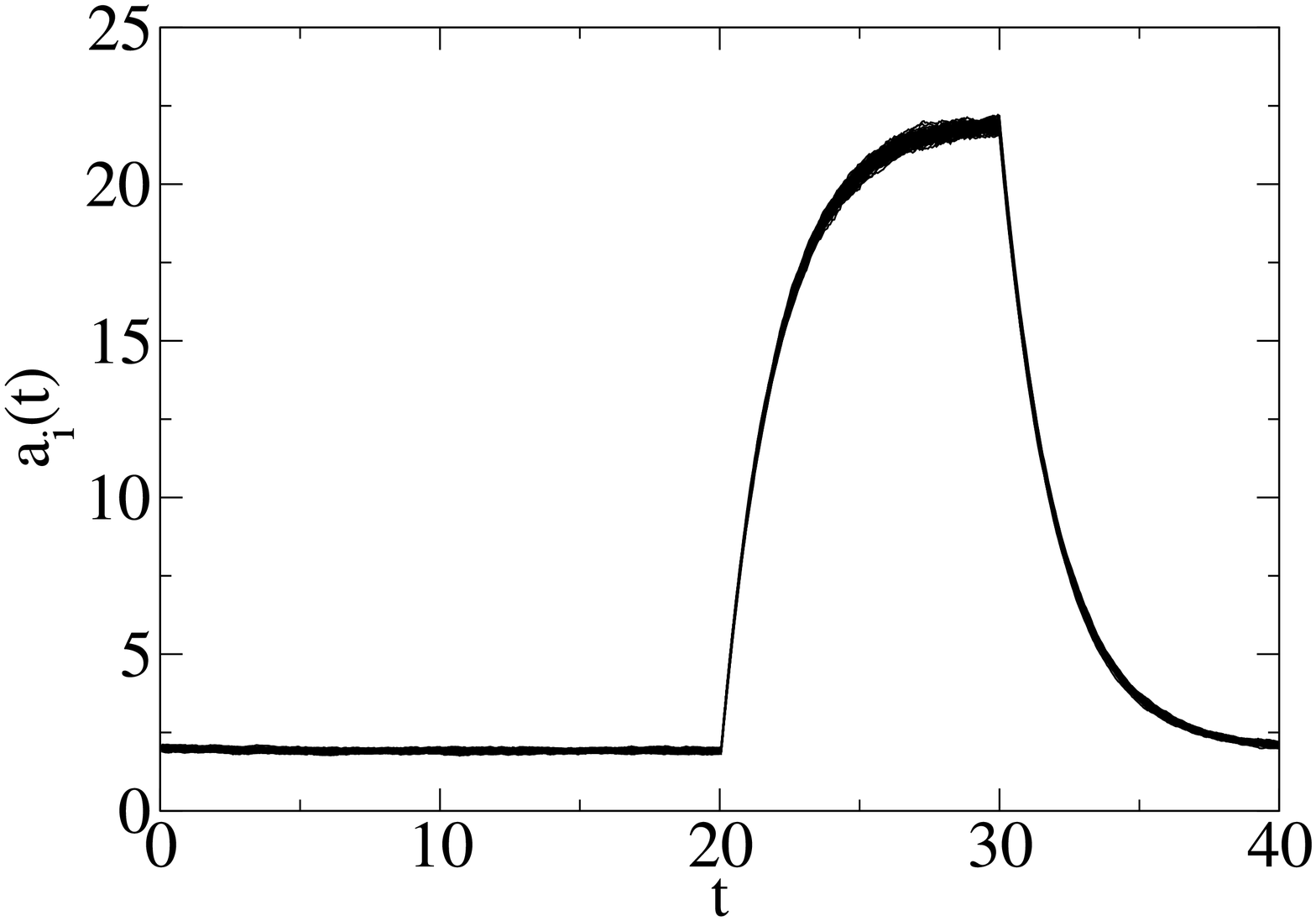}} (b) 
\caption{Response of all-to-all network to correlated input.  $\alpha = 0.5$, $N=100$.  a) the response to correlated input with $\sigma = 100$.  b) the response to a Poisson process with rate $\lambda = 10$.  Note the change in scale between the two plots.  }
\label{fig:switchOff}
\end{center}
\end{figure}

One can see that, even though the theory begins to deviate from the simulations near criticality, we still capture the loss of stability of the active state, even for $N=10$.  This is due to the growth of correlations which negatively feedback on the mean due to the negative second derivative of $f(s)$.  We can use this feature to observe the effect of correlated input directly by adding a term to the Markov process  which provides multiplicative Gaussian noise.  In particular, we add a transition at rate:
\begin{equation}
	\Theta(a_i)  \left ( \sum  a_j \right )^2 \sigma^2 \eta_i
\end{equation}
where $\eta$ is a Gaussian noise source.  The purpose of the step function is to prevent an individual neuron from getting a kick which will drive the activity negative.  Note that we are not using an ``input" term as we have defined it. Because the firing rate function is strictly positive, we cannot use a stimulus such that the mean transition rate is strictly zero.  However, we can use a stimulus such that the source to the correlation function is much stronger than the mean.  For our example, to maximize the effect, we do not use an input to the firing rate function but instead add another transition to the Markov process separate from the firing rate function and the decay transition.  This allows us to source only the correlation function.  Although this is artificial, this transition adds terms to the coupled equations which would approximate those from some input with zero mean and large correlations.
In the following simulations we used $\alpha = 0.5$, $\sigma = 100$ and $N=100$.  The correlated input was given between $t=20$ and $t=30$.  In the absence of external correlated input, these parameters result in the active state being stable, as shown in Figure~\ref{fig:sim100a}.  As can be seen in Figure~\ref{fig:switchOff}, the use of the global noise source results in the ``switch off" behavior predicted by the generalized equations.   If we instead use a Poisson process to provide this external stimulation, one can also see in Figure~\ref{fig:sim100a} that the network responds and then reverts back to the active state.  The reason for the explicit shut off is that the system has an absorbing state to which it is driven.  The more important point is that the correlated input is acting as a source of inhibition whereas the Poisson input serves as an excitatory input.  A linear system or one in which the firing rate function $f(s)$ is such that $f''(s) >0$ will not exhibit this behavior.  We chose this particular example for sourcing $C_{ij}(t)$ in order to separate the effects of sourcing $a_i(t)$ as well.  Given a more complicated noise source, one would need to examine the phase plane or solve the equations, after determining the effects of the noise source on the normal ordered cumulants.


\section{Discussion}

We have demonstrated a formalism for constructing a minimal extension of a rate equation to include correlated activity.  We have explicitly computed the lowest order results of this extension which provide coupled equations for the mean activity, two-point correlations, and linear responses.  These results indicate that correlations can have an important impact on the dynamics of a rate equation, affecting both stability and the structure of bifurcations.  
Our argument relied upon inferring a ``minimal" Markov process.  Our use of the Doi-Peliti path integral formalism guides our assertion that our inferred Markov process is the simplest one compatible with both the rate equations and their interpretation as measuring some stochastic counting process.  Thus, a general extension for any type of rate equation should share the same basic structure that we have described here.  We performed this construction on a Markov process consistent with the Wilson-Cowan equation but our prescription would work equally well with any Markov process.

In keeping with this idea, our results have something in common with other approaches to understanding correlations in neural networks.  El Boustani and Destexhe \cite{dest} attempt to derive a Markov model for the asynchronous irregular states of an underlying neural system and explore the moment hierarchy of that Markov model.  We take the opposite approach, beginning with a presumed set of rate equations and asking what possible restrictions can be placed upon the correlation functions knowing only the dynamics of the rate model.  Their hierarchy is truncated via scaling and finite size, whereas our hierarchy's truncation (and the truncation of the loop expansion) arises through the distance to a bifurcation in the rate equations.  
Ginzburg and Sompolinsky \cite{somp1} propose a simple Markov model and study its moment hierarchy.  For the correlations, they achieve results similar to the tree level of our loop expansion.  An important point of departure is that we consider the recurrent effects the correlations have upon the mean field, which we demonstrate can be sufficiently significant to alter the structure of the bifurcation.

As we predict, our theory breaks down sufficiently close to a bifurcation.  Examining the dynamics near the critical point requires a different form of analysis such as a renormalization argument.  An example  was presented in \cite{bc} where it was argued  that a large class of neural models would exhibit scaling laws near a bifurcation like those of the Directed Percolation model \cite{Janssen2005147}.  The predictions of this scaling coincide with measurements made in cortical slices of ``avalanches" \cite{beggs}.  If criticality is important for neural function \cite{beggs2}, then these sort of approaches will be more important for future work and our rate model extension will be less applicable.  

In contrast, supporting the potential usefulness of our rate model extension is the fact that large neural connectivity suppresses correlations and aids the truncation of the hierarchy, an analogous result to the Ginsburg criterion in equilibrium statistical mechanics.  In addition, we demonstrated that Poisson-like input in general  pushes the system into configurations in which the correlations are suppressed relative to the mean.  All of this suggests that our extension will be at least as applicable as the rate models themselves.

Regarding that applicability, both the Markov process and the rate equations assume a large degree of underlying asynchrony in the network.  The expansion we describe should be appropriate for networks in which a relatively small amount of synchrony at the level of individual neurons is developing.  The coupled correlation function captures this effect.  If the population is being dominated by neuron level synchrony, then the Markov process should no longer hold as a description of the system.  Population level synchrony as captured by the original rate model, however, should have no effect on our analysis.  In other words, there will be correlation effects acting on oscillating states, for example, such as presumably correlation induced modulation of the frequency of the oscillation.  We will demonstrate this in future work.

An important outstanding point is that we have posited this Markov process based on the original interpretation of the Wilson-Cowan equations as dynamical equations for the  probabilistic activity of neurons.  Although our Markov process is the most ``natural" given the transitions in the Wilson-Cowan equations, there is no {\it a priori} reason to suppose that this Markov process reflects the probabilistic dynamics of a physiologically based neural model or of real neurons precisely because there is nothing which  mandates this interpretation.  Per the renormalization analysis of \cite{bc}, measuring scaling laws in cortex will provide a means of identifying classes of models (by identifying the relevant universality class) but this will in no way distinguish between models within the same class.  Distinguishing models within the same class will require the measurement 
of non-universal quantities.  This would likely mean some relatively precision measurements of response functions in cortical activity.

Nonetheless, we feel our approach is a useful starting point for understanding effects beyond mean field.  We have demonstrated a correlation induced loss of stability in an all-to-all network.  This effect should carry over to non-homogeneous solutions such as bump solutions or traveling waves.  Likewise, correlations will modify important aspects of mean field solutions such as dispersion relations.  Our approach enables this dispersion relation to be calculated.  In addition to stability, our equations are a useful starting point towards understanding the wandering of bump solutions.  They also provide a natural means of studying beyond mean field effects of correlation based learning.  A model of visual hallucinations in cortex based on the Turing mechanism has explained many hallucinatory effects (such as the various Kluver form constants).  Since the Turing mechanism is based upon bifurcations, it is an interesting question to what extent the coupling with correlations effects the results of the hallucination analysis.  Our approach may provide this model with a means of explaining further hallucinations not covered by the model in  \cite{bressloff}.

It remains an important question how to connect our Markov  and generalized rate model approach with models of deterministic neurons.  While the formalism admits almost any gain function, there remains the question of connecting this gain function to, for example, the transfer function for some neural model of which the Markov process is some approximation.  This is of course not a question of the analysis of Markov models but of the applicability of rate models as high level descriptions of more detailed neural models.
Answering this question will likely involve a kinetic theory formulation of the neural systems, such as the one pursued in \cite{Hildebrand:2006p4, buice:031118}.  Having derived the generalized equations, it is also now important to explore their further consequences for phenomena such as pattern formation.  
There are also many avenues to extend this model and this approach.  The Markov process can be enlarged to account for synaptic adaptation by adding a synaptic time variable to the neural configuration.  Likewise noise in the transitions themselves, whether spatial or temporal, is easily incorporated into the action.  A reduction of the resulting theory would no longer satisfy the Markov property, although there may be certain natural assumption (such as slow dynamics for the auxiliary field) that could allow one to regain Markovicity with an approximate model.
This would allow us to construct extended Wilson-Cowan equations which incorporate these and other aspects of neural dynamics.
These questions will be explored in future work.

\section{Acknowledgments}
We thank Vipul Periwal for helpful suggestions.  This work was supported by the Intramural Research Program of the NIH/NIDDK.

\appendix

\section{Composite Operator Effective Action and the 2PI equations}
\label{app:2pi}

Here we derive the 2PI equations.  We begin with the generating functional:
\begin{eqnarray}
		Z[ J^{\mu}, K^{\mu\nu} ] &=&  \int {\cal D}\Phi_{\mu} \exp{\left (-\frac{1}{h}\left [S[\Phi_{\mu}] - \int d^dx \,dt\, J^{\mu}(x,t) \Phi_{\mu}(x,t) \right . \right .} \nonumber \\
		&& \left . \left . - \frac{1}{2} \int d^dx\,d^dx'\,dt\,dt'\, \Phi_{\mu}(x,t) K^{\mu\nu}(x,t;x',t') \Phi_{\nu}(x',t')  \right]\right ) 
\end{eqnarray}
where we have introduced a parameter $h=1$ for bookkeeping purposes.
The generalized effective action is given by
\begin{eqnarray}
		\Gamma[a_{\mu}, C_{\mu\nu}] &=& -W[J^{\mu}, K^{\mu\nu} ] + \int d^dx\, dt\, J^{\mu}(x,t) a_{\mu}(x,t)  \\
		&& + \frac{1}{2} \int d^dx\,d^dx'\, dt\,dt'\, \left [ a_{\mu}(x,t) a_{\nu}(x',t') + h C_{\mu\nu}(x,t; x',t')\right]K^{\mu\nu}(x,t;x',t') \nonumber
\end{eqnarray}
with $J^{\mu}$ and $K^{\mu\nu}$ given by
\begin{eqnarray}
		\frac{\delta W[J^{\mu}, K^{\mu\nu}]}{\delta J^{\mu}(x,t)} &=& a_{\mu}(x,t) \\
		\frac{\delta W[J^{\mu}, K^{\mu\nu}]}{\delta K^{\mu\nu}(x,t;x',t')} &=& \frac{1}{2} \left [ a_{\mu}(x,t) a_{\nu}(x',t') + h C_{\mu\nu}(x,t; x',t')\right ]
\end{eqnarray}
The path integral representation is thus
\begin{eqnarray}
		\exp({-\frac{1}{h}\Gamma[a_{\mu}, C_{\mu\nu}] }) &=&  \int {\cal D}\Phi_{\mu} \exp{\left (-\frac{1}{h}\left [S[\Phi_{\mu}] - \int d^dx \,dt\, J^{\mu}(x,t) ( \Phi_{\mu}(x,t) - a_{\mu}(x,t) ) \right . \right .} \nonumber \\
		&& \left .  - \frac{1}{2} \int d^dx\,d^dx'\,dt\,dt'\, (\Phi_{\mu}(x,t)\Phi_{\nu}(x',t') - a_{\mu}(x,t) a_{\nu}(x',t')\right . \nonumber \\
		&& \left . \left . - h C_{\mu\nu}(x,t; x',t'))K^{\mu\nu}(x,t;x',t')   \right]\right ) 
\end{eqnarray}
which we transform to a new variable $\Psi_{\mu}(x,t)=\Phi_\mu(x,t) -a_\mu(x,t)$, set 
\begin{equation}
	J^{\mu} = \frac{\delta \Gamma[a_{\mu}, C_{\mu\nu}]}{\delta a_{\mu}}+\frac{1}{2} \int d^dx'\,dt'\, a_{\nu}(x',t') \left [ K^{\mu\nu}(x,t; x',t') + K^{\mu\nu}(x',t'; x,t)\right ]
\end{equation}
and expand $S[\Psi_\mu+a_\mu]=S[a_\mu]+\int d^dx\, dt\, (L^{\mu}[a_\mu]\Psi_\mu+(1/2)\Psi_\mu L^{\mu \nu}[a_\mu]\Psi_\nu)+V[\Psi_\mu, a_\mu]$
to obtain
\begin{eqnarray}
\exp({-\frac{1}{h}\Gamma[a_{\mu}, C_{\mu\nu}] }) 
&=&  \exp(-\frac{1}{h} S[a_{\mu}]})\int {\cal D}\Psi_{\mu} 
\exp{\left (-\frac{1}{h}\left [ \int d^dx\,dt\, \left(L^{\mu}[a_{\mu}] \Psi_\mu +\frac{1}{2}L^{\mu \nu}[a_\mu]\Psi_\mu\Psi_\nu\right)    \right . \right .   \nonumber \\
& +&\left.\left.  V[\Psi_\mu, a_\mu] - \int d^dx \,dt\,  \Gamma^{\mu,0}[a_{\mu}, C_{\mu\nu}] \Psi_{\mu}(x,t)  \right . \right .   \nonumber \\
&-& \left .  \left . \frac{1}{2} \int d^dx\,d^dx'\,dt\,dt'\, (\Psi_{\mu}(x,t)\Psi_{\nu}(x',t') 
- h C_{\mu\nu}(x,t; x',t'))K^{\mu\nu}(x,t;x',t')   \right]\right ) \nonumber
\end{eqnarray}
where we use 
$$\Gamma^{\mu,0}[a_{\mu}, C_{\mu\nu}]\equiv \frac{\delta \Gamma[a_{\mu}, C_{\mu\nu}]}{\delta a_{\mu}}.$$
We consider the expansion $\Gamma =\Gamma_0+h\Gamma_1+h^2\Gamma_2$, where $\Gamma_2$ contains all terms of order $h^2$ and higher.
Setting $\Gamma_0=S$ gives
\begin{eqnarray}
		\lefteqn{\exp({-\frac{1}{h}\Gamma[a_{\mu}, C_{\mu\nu}] }) =  \exp({-\frac{1}{h} S[a_{\mu}]})
		\int {\cal D}\Psi_{\mu} \exp{\left (-\frac{1}{h}\left [\int d^dx \,dt\,\frac{1}{2}L^{\mu \nu}[a_\mu]\Psi_\mu\Psi_\nu\right . \right .}} \nonumber \\
		&& - \int d^dx \,dt\,  ( h\Gamma_1^{\mu,0}[a_{\mu}, C_{\mu\nu}] + h^2\Gamma_2^{\mu,0}[a_{\mu}, C_{\mu\nu}])\Psi_{\mu}(x,t)   \nonumber \\
		&& \left . \left . - \frac{1}{2} \int d^dx\,d^dx'\,dt\,dt'\, (\Psi_{\mu}(x,t)\Psi_{\nu}(x',t')   - h C_{\mu\nu}(x,t; x',t'))K^{\mu\nu}(x,t;x',t')   + V[\Psi_\mu, a_\mu] \right]\right ) \nonumber 
\end{eqnarray}
We now fix $K^{\mu\nu}$ according to
\begin{equation}
	\frac{\delta \Gamma[a_{\mu}, C_{\mu\nu}]}{\delta C_{\mu\nu}(x,t;x',t')}\equiv  \Gamma^{0,\mu\nu}[a_{\mu}, C_{\mu\nu}] = \frac{1}{2}h K^{\mu\nu}(x,t;x',t')
\end{equation}
which gives
\begin{eqnarray}
		\lefteqn{\exp({-\frac{1}{h}\Gamma[a_{\mu}, C_{\mu\nu}] }) =  \exp\left({-\frac{1}{h} S[a_{\mu}]} - \frac{1}{h}{\rm Tr}\,\Gamma^{0,\mu\nu}[a_{\mu}, C_{\mu\nu}] C_{\mu\nu}\right)
		}\nonumber \\
		&&\int {\cal D}\Psi_{\mu} \exp\left (-\frac{1}{h}\left [\int d^dx \,dt\,\frac{1}{2}L^{\mu \nu}[a_\mu]\Psi_\mu\Psi_\nu -    ( h\Gamma_1^{\mu,0}[a_{\mu}, C_{\mu\nu}] + h^2\Gamma_2^{\mu,0}[a_{\mu}, C_{\mu\nu}])\Psi_{\mu}(x,t)  \right . \right .\nonumber \\
				&& \left . \left . - \frac{1}{2} \int d^dx\,d^dx'\,dt\,dt'\, \Psi_{\mu}(x,t)\Psi_{\nu}(x',t') \frac{1}{h}\frac{\delta \Gamma[a_{\mu}, C_{\mu\nu}]}{\delta C_{\mu\nu}} + V[\Psi_\mu, a_\mu]    \right]\right ) 
				\label{eq:preC}
\end{eqnarray}
where
\begin{equation}
	{\rm Tr}\,\Gamma^{0,\mu\nu}[a_{\mu}, C_{\mu\nu}] C_{\mu\nu}= \int d^dx d^dx' dt dt' \frac{\delta \Gamma[a_{\mu}, C_{\mu\nu}]}{\delta C_{\mu\nu}(x,t;x',t')} C_{\mu\nu}
\end{equation}
We need to extract the order $h$ contributions from the functional integral.  We expect the effect of $\Gamma^{0,\mu \nu}$ is to replace $L^{\mu \nu}$ with the full inverse two point function $(C^{-1})^{\mu \nu}$.  This in turn will affect the normalization of the integral.  Because of this we expect the order $h$ contribution to the effective action to be:
\begin{equation}
	\Gamma_1[a_\mu, C_{\mu\nu}] =  \frac{1}{2}  {\rm Tr} \ln \left (C^{-1} \right)^{\mu\nu} + \frac{1}{2} {\rm Tr} L^{\mu \nu}[a_\mu] C_{\mu\nu} + {\rm constant}
\end{equation}
Substituting this into expression~(\ref{eq:preC}) gives us
\begin{eqnarray}
		\exp({-\frac{1}{h}\Gamma[a_{\mu}, C_{\mu\nu}] }) &=&  \exp\left({-\frac{1}{h} S[a_{\mu}]} - \frac{1}{2} {\rm Tr} L^{\mu \nu}[a_\mu] C_{\mu\nu} + \frac{1}{2} {\rm Tr} {\bf 1}_{\mu \nu}   - h{\rm Tr}\,\Gamma_2^{0,\mu\nu}[a_{\mu}, C_{\mu\nu}] C_{\mu\nu}\right) \nonumber \\
		&\times& \int {\cal D}\Psi_{\mu} \exp{\left (-\frac{1}{h}\left [\frac{1}{2}(C^{-1})^{\mu \nu}\Psi_\mu\Psi_\nu \right . \right .} \nonumber \\
		&&+ V[\Psi_\mu, a_\mu]   -  \int d^dx \,dt\,  ( h\Gamma_1^{\mu,0}[a_{\mu}, C_{\mu\nu}] + h^2\Gamma_2^{\mu,0}[a_{\mu}, C_{\mu\nu}])\Psi_{\mu}(x,t)  \nonumber \\
				&& \left . \left . - \frac{1}{2} \int d^dx\,d^dx'\,dt\,dt'\, \Psi_{\mu}(x,t)\Psi_{\nu}(x',t')h \frac{\delta  \Gamma_2[a_{\mu}, C_{\mu\nu}]}{\delta C_{\mu\nu}} \right]\right ) 
				\label{eq:postC}
\end{eqnarray}
We can extract the normalization of the functional integral using
$
	\frac{1}{2} \ln \det C_{\mu \nu} = - \frac{1}{2} \ln \det (C^{-1})^{\mu \nu}
$
and the identity
$
	\ln \det A = {\rm Tr} \ln A
$
to obtain
\begin{eqnarray}
		\exp({-\frac{1}{h}\Gamma[a_{\mu}, C_{\mu\nu}] }) &=&  \exp\left({-\frac{1}{h} S[a_{\mu}]} - \frac{1}{2} {\rm Tr} L^{\mu \nu}[a_\mu] C_{\mu\nu} -\frac{1}{2}  {\rm Tr} \ln \left (C^{-1} \right)^{\mu\nu} + \frac{1}{2} {\rm Tr} {\bf 1}_{\mu \nu} - h{\rm Tr}\,\Gamma_2^{0,\mu\nu}[a_{\mu}, C_{\mu\nu}] C_{\mu\nu}\right) \nonumber \\
		&\times& \left (\sqrt{\det C_{\mu \nu}} \right) \int {\cal D}\Psi_{\mu} \exp{\left (-\frac{1}{h}\left [\frac{1}{2}(C^{-1})^{\mu \nu}\Psi_\mu\Psi_\nu \right . \right .} \nonumber \\
		&&+ V[\Psi_\mu, a_\mu]   -  \int d^dx \,dt\,  ( h\Gamma_1^{\mu,0}[a_{\mu}, C_{\mu\nu}] + h^2\Gamma_2^{\mu,0}[a_{\mu}, C_{\mu\nu}])\Psi_{\mu}(x,t)  \nonumber \\
				&& \left . \left . - \frac{1}{2} \int d^dx\,d^dx'\,dt\,dt'\, \Psi_{\mu}(x,t)\Psi_{\nu}(x',t')h \frac{\delta  \Gamma_2[a_{\mu}, C_{\mu\nu}]}{\delta C_{\mu\nu}} \right]\right ) 
				\label{eq:finalC}
\end{eqnarray}
The factor of the determinant serves as a normalization for the functional integral, which is now in a form that will only contribute to the effective action at order $h^2$.
Thus we have
\begin{eqnarray}
		\Gamma[a_{\mu}, C_{\mu\nu}] &=& S[a_{\mu}] + \frac{1}{2} h {\rm Tr} \ln \left (C^{-1} \right)^{\mu\nu} + \frac{1}{2}h {\rm Tr} L^{\mu\nu}[a_\mu]C_{\mu\nu} \nonumber \\
		&& + \Gamma_2 [a_{\mu}, C_{\mu\nu}] - \frac{1}{2} h{\rm Tr}\, {\bf 1}_{\mu\nu} \label{eq:effaction}
\end{eqnarray}
	


Now we can calculate the equations of motion to a given loop order from equations~(\ref{eq:phieom}) and (\ref{eq:geom}).
	Using equation~(\ref{eq:phieom}), the equations of motion for the mean field are
	\begin{eqnarray}
		\frac{\delta S[a_{\mu}]}{\delta a_{\mu}} + \frac{1}{2}h\frac{\delta}{\delta a_{\mu}}{\rm Tr} L^{\mu\nu}[a_\mu]C_{\mu\nu}   + \frac{\delta \Gamma_2[a_{\mu}, C_{\mu\nu}]}{\delta a_{\mu}} = 0
		\label{eq:phieom1}
	\end{eqnarray}
	The equations for $C^{\mu\nu}$ are
	\begin{eqnarray}
		-\frac{1}{2}h (C^{-1})^{\mu\nu} + \frac{1}{2}h L^{\mu\nu}[a_\mu] + \frac{\delta \Gamma_2[a_{\mu}, C_{\mu\nu}]}{\delta C_{\mu\nu}} = 0
	\end{eqnarray}
which we can invert to get
	\begin{equation}
		L^{\mu\nu}[a_\mu]C_{\mu\nu} + \frac{2}{h}\frac{\delta \Gamma_2[a_{\mu}, C_{\mu\nu}]}{\delta C_{\mu\nu}}  C_{\mu\nu} = \delta_{\mu\nu}
		\label{eq:geom1}
	\end{equation}
	In particular, if we ignore loop corrections (i.e. only consider first order in $h$, recalling that $\Gamma_2$ is $O(h^2)$), we get 
	\begin{equation}
		C_{\mu\nu} = L^{-1}_{\mu\nu}[a_\mu]
	\end{equation}
	where $L^{-1}_{\mu\nu}[a_\mu]$ is the inverse of $L^{\mu \nu}[a_\mu]$.  In the absence of interactions, $L^{-1}_{\mu \nu}$ is the two-point function, as expected.

We can use the loop expansion to draw some conclusions about the applicability of perturbation theory.  Since $\Gamma_2[a_{\mu}, C_{\mu\nu}]$ is second order and is the sum of vacuum two particle irreducible graphs, every graph contributing to it must be at least of two loop order.  Every internal line represents a factor of $C_{\mu\nu}$ and so each graph contributing to $\Gamma_2[a_{\mu}, C_{\mu\nu}]$ must have at least two factors of $C_{\mu\nu}$, each of which will either be equal to $0$, or be attenuated (in steady state) by the same exponents which attenuate the magnitude of $C(x,y,t)$ away from a bifurcation, according to equations~(\ref{eq:flucdis0}-\ref{eq:flucdis}).  Thus the argument that $C_{\mu\nu}$ is small away from the critical point extends to \emph{every term in the expansion for the generalized equations}.

The caveat here is that there is a class of diagrams which couple the lowest order expression for a given moment to the mean field.  Although these graphs are suppressed by the distance to criticality, each of these is of the same order.  We are assisted by two facts.  The first is that the source terms for each of these moments at lowest order will be proportional to derivatives of the firing rate function.  If $f(s)$ is sufficiently smooth, this will suppress higher order contributions.  In addition, each coupling will go as an additional factor of $N_m^{-1}$ where $N_m$ was defined in section~\ref{sec:crit} as the smallest number of inputs to any given neuron.  Thus the connectivity in cortex will serve to ``average out" sources to the mean from higher moments.  This will be the case as long as we can bound the total input to any given neuron.

\section{Tree level equations of motion}
\label{app:tree}
In order to calculate the expansion for the equations of motion, we need to compute the value of both $L^{\mu \nu}$ and $\Gamma_2$.  We compute the lowest order correction here.
	
	First we find the intermediate results (which give us the classical equations of motion for $a$ and $\tilde{a}$):
	\begin{eqnarray}
		\lefteqn{ L^{1}[a_{\mu}]\equiv \frac{\delta S[a_{\mu}]}{\delta \tilde{a}(x,t)}  = \left ( \partial_t + \alpha \right) a(x,t) - f \left ( w \star \left [ \tilde{a}(x,t) a(x,t) + a(x,t) \right ] \right)} \nonumber \\ 
		&&- \int d^dx'' \, \tilde{a}(x'',t) f^{(1)}\left ( w \star \left [ \tilde{a}(x'',t) a(x'',t) + a(x'',t) \right ] \right) w(x''-x) a(x,t) \label{eq:classical-phi}\\
		\lefteqn{L^{-1}[a_{\mu}]\equiv\frac{\delta S[a_{\mu}]}{\delta a(x,t)}  = \left ( -\partial_t + \alpha \right) \tilde{a}(x,t)} \\
		&&- \int d^dx''\,\tilde{a}(x'',t)f^{(1)}\left ( w \star \left [ \tilde{a}(x'',t) a(x'',t) + a(x'',t) \right ] \right)w(x'' - x) \left [\tilde{a}(x,t) + 1 \right] \nonumber 
		\label{eq:classical-phi-tilde}
	\end{eqnarray}
	from which follows
	\begin{eqnarray}
		L^{-1,-1}[a_{\mu}](x,t; x',t') &=& - \left [ f^{(1)}(x,t) w(x-x') a(x',t') + f^{(1)}(x',t') w(x'-x) a(x,t) \right ] \delta(t-t') \nonumber \\
		&& - \int d^dx''\, \tilde{a}(x'',t) f^{(2)}(x'',t) w(x'' - x) a(x,t) w(x'' - x') a(x',t)\delta(t - t') \\
		L^{-1,1}[a_{\mu}](x,t; x',t') &=& \left ( \partial_t + \alpha \right) \delta(x-x') \delta(t-t') - f^{(1)}(x,t)w(x-x') \left[\tilde{a}(x',t') + 1 \right] \delta(t-t')	\nonumber \\
		&& -\int d^dx''\,  \tilde{a}(x'',t) f^{(1)}(x'',t'') w(x''-x) \delta(x-x')\delta(t-t') \nonumber \\
		&& -\int d^dx''\,  \tilde{a}(x'',t) f^{(2)}(x'',t'') w(x''-x) a(x,t)w(x''-x') \left [ \tilde{a}(x',t') + 1\right] \delta(t-t') \nonumber \\
		\\
		L^{1,-1}[a_{\mu}](x,t; x',t') &=& \left (- \partial_t + \alpha \right) \delta(x-x') \delta(t-t') - f^{(1)}(x',t)w(x'-x) \left[\tilde{a}(x,t) + 1 \right] \delta(t-t')	\nonumber \\
		&& -\int d^dx''\,  \tilde{a}(x'',t) f^{(1)}(x'',t'') w(x''-x) \delta(x-x')\delta(t-t') \nonumber \\
		&& -\int d^dx''\,  \tilde{a}(x'',t) f^{(2)}(x'',t'') w(x''-x') a(x',t)w(x''-x) \left [ \tilde{a}(x,t) + 1\right] \delta(t-t') \nonumber \\
		\\
		L^{1,1}[a_{\mu}](x,t; x',t') &=& - \int d^dx'' \tilde{a}(x'',t) f^{(2)}(x'',t)w(x''-x)w(x''-x') \left[ \tilde{a}(x,t) + 1\right] \left[ \tilde{a}(x',t) + 1\right] \delta(t-t') \nonumber \\
	\end{eqnarray}
	The terms $f^{(n)}(x,t)$ indicate the $n$th derivative of $f$.  Note that we have suppressed the argument, so that $f^{(n)}(x,t) = f^{(n)}\left ( w \star \left [ \tilde{a}(x,t) a(x,t) + a(x,t) \right ] \right)$
	
	
	We can now write down the equations of motion from (\ref{eq:geom1}), minus the loop corrections.  The first ``diagonal" equation (for $(-1,-1)$) is:
	\begin{eqnarray}
		\left ( \partial_t + \alpha \right) C_{1,-1}(x,t; x_0,t_0) - \int d^dx'\, f^{(1)}(x,t)w(x-x') \left[\tilde{a}(x',t') + 1 \right] C_{1,-1}(x',t; x_0,t_0) \nonumber \\
		-\int d^dx''\,  \tilde{a}(x'',t) f^{(1)}(x'',t'') w(x''-x) C_{1,-1}(x,t; x_0,t_0)  \nonumber \\
		 -\int d^dx''\,d^dx'\,  \tilde{a}(x'',t) f^{(2)}(x'',t'') w(x''-x) a(x,t)w(x''-x') \left [ \tilde{a}(x',t') + 1\right] C_{1,-1}(x',t; x_0,t_0)  \nonumber \\
		 - \int d^dx'\,\left [ f^{(1)}(x,t) w(x-x') a(x',t') + f^{(1)}(x',t') w(x'-x) a(x,t) \right ] C_{-1,-1}(x',t;x_0,t_0)  \nonumber \\
		 - \int d^dx''\,d^dx'\, \tilde{a}(x'',t) f^{(2)}(x'',t) w(x'' - x) a(x,t) w(x'' - x') a(x',t)C_{-1,-1}(x',t;x_0,t_0)\nonumber \\
		 = \delta(x-x_0) \delta(t-t_0) \nonumber \\ \label{eq:diag1}
	\end{eqnarray}
	The second ``diagonal" equation (for $11$):
	\begin{eqnarray}
		\left (- \partial_t + \alpha \right)C_{-1,1}(x,t;x_0,t_0) - \int d^dx'\,f^{(1)}(x',t)w(x'-x) \left[\tilde{a}(x,t) + 1 \right] C_{-1,1}(x',t;x_0,t_0)\nonumber \\
		-\int d^dx''\,  \tilde{a}(x'',t) f^{(1)}(x'',t'') w(x''-x)C_{-1,1}(x,t;x_0,t_0)  \nonumber\\
		 -\int d^dx''\, d^dx'\, \tilde{a}(x'',t) f^{(2)}(x'',t'') w(x''-x') a(x',t)w(x''-x) \left [ \tilde{a}(x,t) + 1\right] C_{-1,1}(x',t;x_0,t_0) \nonumber\\
		 - \int d^dx'' \,d^dx' \,\tilde{a}(x'',t) f^{(2)}(x'',t)w(x''-x)w(x''-x') \left[ \tilde{a}(x,t) + 1\right] \left[ \tilde{a}(x',t) + 1\right]  C_{11}(x',t;x_0,t_0) \nonumber \\
		 = \delta(x-x_0)\delta(t-t_0) \nonumber \\  \label{eq:diag2}
	\end{eqnarray}
	The ``off-diagonal" equations are (starting with $-1,1$):
	\begin{eqnarray}
		\left ( \partial_t + \alpha \right) C_{11}(x,t; x_0,t_0) - \int d^dx'\, f^{(1)}(x,t)w(x-x') \left[\tilde{a}(x',t') + 1 \right] C_{11}(x',t; x_0,t_0) &&\\
		-\int d^dx''\,  \tilde{a}(x'',t) f^{(1)}(x'',t'') w(x''-x) C_{11}(x,t; x_0,t_0) && \nonumber\\
		 -\int d^dx''\,d^dx'\,  \tilde{a}(x'',t) f^{(2)}(x'',t'') w(x''-x) a(x,t)w(x''-x') \left [ \tilde{a}(x',t') + 1\right] C_{11}(x',t; x_0,t_0) && \nonumber \\
		 - \int d^dx'\,\left [ f^{(1)}(x,t) w(x-x') a(x',t') + f^{(1)}(x',t') w(x'-x) a(x,t) \right ] C_{-1,1}(x',t;x_0,t_0) && \nonumber \\
		 - \int d^dx''\,d^dx'\, \tilde{a}(x'',t) f^{(2)}(x'',t) w(x'' - x) a(x,t) w(x'' - x') a(x',t)C_{-1,1}(x',t;x_0,t_0)&=& 0 \nonumber
	\end{eqnarray}
	and the other ($1,-1$):
	\begin{eqnarray}
		\left (- \partial_t + \alpha \right)C_{-1,-1}(x,t;x_0,t_0) - \int d^dx'\,f^{(1)}(x',t)w(x'-x) \left[\tilde{a}(x,t) + 1 \right] C_{-1,-1}(x',t;x_0,t_0) \\
		-\int d^dx''\,  \tilde{a}(x'',t) f^{(1)}(x'',t'') w(x''-x)C_{-1,-1}(x,t;x_0,t_0)  \nonumber \\
		 -\int d^dx''\, d^dx'\, \tilde{a}(x'',t) f^{(2)}(x'',t'') w(x''-x') a(x',t)w(x''-x) \left [ \tilde{a}(x,t) + 1\right] C_{-1,-1}(x',t;x_0,t_0)  \nonumber\\
		 - \int d^dx'' \,d^dx' \,\tilde{a}(x'',t) f^{(2)}(x'',t)w(x''-x)w(x''-x') \left[ \tilde{a}(x,t) + 1\right] \left[ \tilde{a}(x',t) + 1\right]  C_{1,-1}(x',t;x_0,t_0) &=&0 \nonumber
	\end{eqnarray}
	
	The ``mean field" portion of the equations of motion~(\ref{eq:phieom1}) are obtained from equations~(\ref{eq:classical-phi}) and (\ref{eq:classical-phi-tilde}) (by setting the LHS to zero).  The remainder of the equations of motion are ``classical" terms dependent on the correlation functions, and loop corrections.  The latter are given by the term $\frac{1}{2}h\frac{\delta}{\delta a_{\mu}}{\rm Tr} L^{\mu\nu}[a_\mu]C_{\mu\nu}$ in equation~(\ref{eq:phieom1}).  The term in the trace is, of course, the sum of the LHS of equations~\ref{eq:diag1} and \ref{eq:diag2}. 
	
	We can simplify the equations for the mean field by realizing that any term involving $C_{-1,1}$ or $C_{1,-1}$ can be ignored because they will only appear in the form $C_{-1,1}(x', t; x, t)$, i.e. at equal initial and final times.  These will be zero.  This can be seen as either the ``initial condition" for the linear response terms or as a manifestation of the ``$\epsilon(0)$" problem in quantum field theory.\cite{zinnjustin} 
	
	Furthermore, we can use some results from the full theory.  In particular, we have
	\begin{eqnarray}
	  	C_{-1,-1}(x,t; x',t') &=& 0\\
		\tilde{a}(x,t) &=& 0 
	\end{eqnarray}
	because causality enforces that $\tilde{\varphi}$ operators can't contract with anything ``in the past".

The equation for $a(x,t)$ is then:
	\begin{eqnarray}
		\left ( \partial_t + \alpha \right) a(x,t) - f \left ( w \star  a(x,t)\right) && \nonumber \\
		- \int d^dx' \, d^dx'' \, f^{(2)}(x,t)w(x-x'')w(x-x')  C_{11}(x',t;x'',t) &=&  0
	\end{eqnarray}
	Applying these same simplifications to the equations for $C_{\mu\nu}$, we get:
	\begin{eqnarray}
		\left ( \partial_t + \alpha \right) C_{1,-1}(x,t; x_0,t_0) &-& \int d^dx'\, f^{(1)} \left ( w \star  a(x,t)\right)w(x-x') C_{1,-1}(x',t; x_0,t_0) \nonumber \\
		&=& \delta(x-x_0) \delta(t-t_0) \label{eq:flucdis0} \\
		\left (- \partial_t + \alpha \right)C_{-1,1}(x,t;x_0,t_0) &-& \int d^dx'\,f^{(1)} \left ( w \star  a(x',t)\right)w(x'-x) C_{-1,1}(x',t;x_0,t_0)  \nonumber \\
		&=& \delta(x-x_0)\delta(t-t_0) \\
		\left ( \partial_t + \alpha \right) C_{11}(x,t; x_0,t_0) &-& \int d^dx'\, f^{(1)} \left ( w \star  a(x,t)\right)w(x-x') C_{11}(x',t; x_0,t_0)  \nonumber \\
		 - \int d^dx'\,\left [ f^{(1)} \left ( w \star  a(x,t)\right) w(x-x') a(x',t') \right . &+& \left . f^{(1)} \left ( w \star  a(x',t)\right) w(x'-x) a(x,t) \right ] C_{-1,1}(x',t;x_0,t_0) \nonumber \\
		 &=&0 \label{eq:flucdis} 
	\end{eqnarray}
	together with the conditions
	\begin{eqnarray}
		C_{11}( x,t; x',t') &=& C_{11}( x',t'; x,t) \\
		C_{-1,1}( x,t; x',t') &=& C_{1,-1}( x',t'; x,t)
	\end{eqnarray}

\bibliography{GenWC13.bib}
\bibliographystyle{apalike}

\end{document}